\documentclass[12pt,showpacs,showkeys,prb]{revtex4}

\usepackage{amssymb}
\usepackage{amsmath}
\usepackage{graphicx}
\usepackage{amsbsy}

\newcommand{\be}{\begin{equation}}
\newcommand{\ee}{\end{equation}}
\newcommand{\bq}{\begin{eqnarray}}
\newcommand{\eq}{\end{eqnarray}}
\newcommand{\bqn}{\begin{eqnarray*}}
\newcommand{\eqn}{\end{eqnarray*}}

\newcommand{\avdel}{\langle\delta\rangle}
\newcommand{\avdels}{\langle\delta^2\rangle}

\newcommand{\mom}{\rho}
\newcommand{\pot}{\phi}
\newcommand{\sg}{\sqrt{\gamma}}
\newcommand{\ex}{^{\rm ex}}
\newcommand{\pa}{{(0)}}
\newcommand{\one}{^{(1)}}
\newcommand{\two}{^{(2)}}
\newcommand{\mono}{_0}
\newcommand{\kT}{k_{\rm B}T}

\begin{document}
\title{Phase behavior of weakly polydisperse sticky hard spheres:
Perturbation theory for the Percus-Yevick solution}

\author{Riccardo Fantoni}
\email{rfantoni@unive.it}
\author{Domenico Gazzillo}
\email{gazzillo@unive.it}
\author{Achille Giacometti}
\email{achille@unive.it}
\affiliation{Istituto Nazionale per la Fisica della Materia and
Dipartimento di Chimica Fisica, Universit\`a di Venezia, S. Marta DD
2137, I-30123 Venezia, Italy} 
\author{Peter Sollich}
\email{peter.sollich@kcl.ac.uk}
\affiliation{King's College London, Department of Mathematics, Strand,
London WC2R 2LS, U.K.} 

\date{\today}

\begin{abstract}
\noindent We study the effects of size polydispersity on the
gas-liquid phase behaviour of mixtures of
sticky hard spheres.
To achieve this, the system of coupled quadratic equations for the contact
values of the partial cavity functions of the Percus-Yevick solution
is solved
within a perturbation expansion in the polydispersity, i.e.\ the
normalized width of the size distribution.
This allows us to make predictions for various thermodynamic quantities
which can be tested against numerical simulations and experiments.
In particular, we determine the leading-order effects of size
polydispersity on the cloud curve delimiting the region of two-phase
coexistence and on the associated shadow curve; we also study the extent
of size fractionation between the coexisting phases. Different choices
for the size-dependence of the adhesion strengths are examined
carefully; the Asakura-Oosawa model of a mixture of polydisperse
colloids and small polymers is studied as a specific example.
\end{abstract}

\maketitle

\section{Introduction}
In the context of soft matter, a number of systems are known
to display a combination of a very steep repulsion and
a short range attraction. This includes for instance
polymer-coated colloids \cite{Lowen94,Nagele96}, globular proteins 
\cite{Lyklema5} and microemulsions \cite{Lyklema1}.
In spite of the notable differences in the details of the
interactions among these systems, most of the common essential
features are captured by a paradigmatic model  
known as the adhesive or sticky hard sphere model.
Sticky hard spheres are impenetrable particles of diameters 
$\{\sigma_i\}$ with an adhesive surface. The simplest
way of describing the adhesion properties, in the framework of atomic fluids,
was originally proposed by Baxter \cite{Baxter68} 
in terms of a potential where energy and length scales
were combined into a single parameter, thus defining the so-called sticky hard sphere (SHS) potential.
Baxter showed that for this model the Ornstein-Zernicke integral equation determining the
correlation functions in the liquid state 
admitted an analytic solution within the Percus-Yevick
(PY) approximation. Together with his collaborators, he predicted
from this solution (via both the 
compressibility and the energy routes of liquid state theory) that the
model displays a gas-liquid transition \cite{Baxter71,Watts71}.
This PY solution was soon extended to mixtures
\cite{Baxter70,Barboy75,Perram75,Barboy79} and has since found
a number of interesting applications in the area of colloidal suspensions 
\cite{Robertus89,Chen94,Lowen94,Nagele96,Piazza98,Prinsen04}. When 
studying the phase behavior of such fluids an important issue 
to deal with is the fact that colloidal particles are generally not
identical but may have different
characteristics (size, charge, chemical species etc). Often, the
distribution of the relevant parameter is effectively
continuous and the fluid is then referred to as polydisperse.
We will focus in this paper on {\em size polydispersity}, i.e.\ a
fluid with a distribution of particle diameters. (A small
degree of size polydispersity is in fact 
required to resolve thermodynamic pathologies which occur
in the case of strictly equal-sized, i.e.\ monodisperse, sticky hard
spheres \cite{Stell91}.) The particle size distribution is fixed when
the particles are synthesized. Thereafter, only the overall density
can be modified by adding or removing solvent, while keeping constant all
ratios of densities of particles of different size; this traces out a
so-called ``dilution line'' in the phase diagram.

Given the success of the PY closure for the monodisperse SHS model, it
is natural to try to extend it to the polydisperse
case. Unfortunately, the PY approximation is tractable only for
mixtures of a small number of particle species: the case of a binary
mixture can be solved analytically \cite{Barboy79}, and for mixtures
with a limited number of components (10 or fewer) a numerical solution
is feasible \cite{Robertus89}. The polydisperse case requires one to
keep track of an effectively infinite number of particles species, one
for each size, and cannot be tackled directly. An alternative, which
we have explored in past work, is to use simpler 
integral equation theories such as the modified Mean
Spherical Approximation 
(mMSA or C0). Between this and the Percus Yevick (PY) approximation
\cite{Baxter68} lie a set of increasingly accurate approximations,
denoted as C$n$ with $n=1,2,\ldots$ They are based
on a density expansion of the direct correlation function
outside the hard core and can be shown
to improve, order by order, the various virial coefficients \cite{Gazzillo04}.
These C$n$ approximations {\em can} be extended to the polydisperse
case with relative ease, provided a particular factorization
holds for the matrices appearing in the solution of Baxter's
equations. This has allowed us to perform a comprehensive analysis of
polydispersity effects on the gas-liquid phase separation \cite{Fantoni05a}. 

The tractability of the C$n$ approximations for the polydisperse SHS
model does, however, come at the price of lower accuracy. Indeed, 
for the monodisperse case accurate Monte Carlo
simulation data recently published by Miller and Frenkel 
\cite{Miller03,Miller04a,Miller04b} 
show that the equation of state of the fluid lies very close to the one 
derived from the energy route of the PY closure. Both the C0 and
C1 approximations, on the other hand, yield precise results only
within a rather limited 
region of the phase diagram, corresponding to high temperatures or
low densities \cite{Gazzillo04}; see Fig.~\ref{fig:oceos} below.

The above considerations show that another attack on the PY closure
for polydisperse SHS fluids is worthwhile in order to get accurate
predictions for the gas-liquid phase
behavior. Rather than trying to tackle the most general case of a fluid
with a potentially wide distribution of particle sizes, which for now
remains out of reach, we exploit the
idea of Evans \cite{Evans01} to treat size polydispersity as a
perturbation to the monodisperse phase behavior.
For this method to apply, the size distribution only has to be
sufficiently narrow but its shape is otherwise arbitrary.
Our approach is also of sufficient generality to consider arbitrary
dependences of the adhesion strengths on the particle sizes,
including those considered in previous work on
the C$n$ approximations \cite{Fantoni05a,Fantoni05b}.
Throughout, we consider gas-liquid phase coexistence only. It has been
argued~\cite{Sear99d} that even in the presence of polydispersity this
is metastable with respect to phase separation into a colloidal gas and
solid. However, the latter may be unobservable on realistic timescales
when formation of the polydisperse solid is hindered by large
nucleation barriers~\cite{AueFre01} or an intervening kinetic glass
transition~\cite{PusVan87}; the gas-liquid phase splits we calculate
will then control the physically observable behaviour. Even where the
kinetics does allow formation of solid phases, the metastable gas-liquid
phase behaviour can play a role, e.g.\ in determining phase
ordering pathways~\cite{PooRenEvaFaiCatPus99}.

The paper is organized as follows. In section \ref{sec:model} we
describe the polydisperse SHS model and discuss various routes for
predicting the thermodynamics of this system, comparing their accuracy
for the better understood monodisperse case. In the polydisperse
setting one needs to model how the strength of the adhesion between
two particles depends on their size; we discuss some possible choices for
this in section \ref{sec:coeff}. Section \ref{sec:core} describes our
perturbation expansion of the PY closure for the weakly polydisperse
SHS model. We first define the perturbation
expansion of the free energy used by Evans (Sec.\ \ref{sec:evans})
and summarize the relevant consequences 
for two-phase coexistence and the attendant size fractionation
effects. The basic equations
that need to be solved in order to determine thermodynamic properties
within the PY approximation are then described
and solved perturbatively (Sec.\ \ref{sec:py}), while Sec.\
\ref{sec:thermodynamics} 
derives from this, via the energy route, the excess Helmholtz free
energy. In section \ref{sec:phases} we evaluate numerically the
consequences of polydispersity for two-phase coexistence and
fractionation for a number of example scenarios, and compare with
alternative approximation schemes.
Section \ref{sec:conclusions} gives concluding remarks.


\section{The SHS model}
\label{sec:model}
The $p$-component SHS
mixture model is made up of Hard Spheres (HS) of different diameters
$\sigma_i$, $i=1,2,\ldots,p$, interacting through a particular pair
potential defined via the following limit procedure. One starts
with a pair interaction potential $\phi_{ij}(r)$ with a hard core
extending out to distance $r=\sigma _{ij}=(\sigma_i+\sigma_j)/2$,
followed by a square well potential of width $R_{ij}-\sigma _{ij}$:
\bq
\phi_{ij}(r)=\left\{ 
\begin{array}{ll}
+\infty & 0<r<\sigma _{ij}~, \\ 
\displaystyle-\ln \left(\frac{1}{12\tau_{ij}}\frac{R_{ij}}{R_{ij}
-\sigma_{ij}}\right) & \sigma _{ij}\leq r\leq R_{ij}~, \\ 
0 & r>R_{ij}~,
\end{array}
\right.  \label{SW}
\eq
Here 
the dimensionless parameter 
\bq \label{tauij}
\frac{1}{\tau_{ij}}=\frac{\epsilon _{ij}}{\tau }\geq 0
\eq
measures the surface adhesion strength or \textquotedblleft
stickiness\textquotedblright\ between particles of species $i$ and
$j$. In Eq.~(\ref{tauij}) the reduced temperature $\tau$ is an
unspecified increasing function of the physical temperature $T$; the
coefficients
$\epsilon_{ij}$ specify how stickiness depends on which particle
species are in contact and are discussed more fully in the next
section. The procedure which defines
the SHS model then consists in taking the ``sticky limit'' $R_{ij}
\rightarrow \sigma _{ij}$.
The logarithm in the initial square well
potential (\ref{SW}) is chosen such as to give a simple expression for the
Boltzmann factor $\exp[-\phi_{ij}(r)]$, which reduces to a
combination of a Heaviside step 
function and a Dirac delta function in the sticky limit. Here and in
the following we measure all energies in units of $\kT$, to simplify
the notation.

A fully polydisperse system is obtained from the above discrete
mixture by replacing
the molar fractions $x_i=N_i/N$, where $N_i$ is the number of
particles of species $i$ and $N$ the total number of particles,
with a normalized size
distribution function $p(\sigma)$:
\bqn
x_i\longrightarrow p(\sigma)\,d\sigma~.
\eqn
Here $p(\sigma)\,d\sigma$ is the fraction of spheres with diameter in
the interval $(\sigma,\sigma+d\sigma)$. Similarly, given a quantity $a_i$ that
depends on the species index one replaces
\bqn
a_i&\longrightarrow& a(\sigma)~,\\
\langle a\rangle=\sum_{i}x_ia_i&\longrightarrow& \int_0^\infty
a(\sigma)p(\sigma)\,d\sigma~.
\eqn

\begin{figure}[ht!]
\begin{center}
\includegraphics[width=7cm]{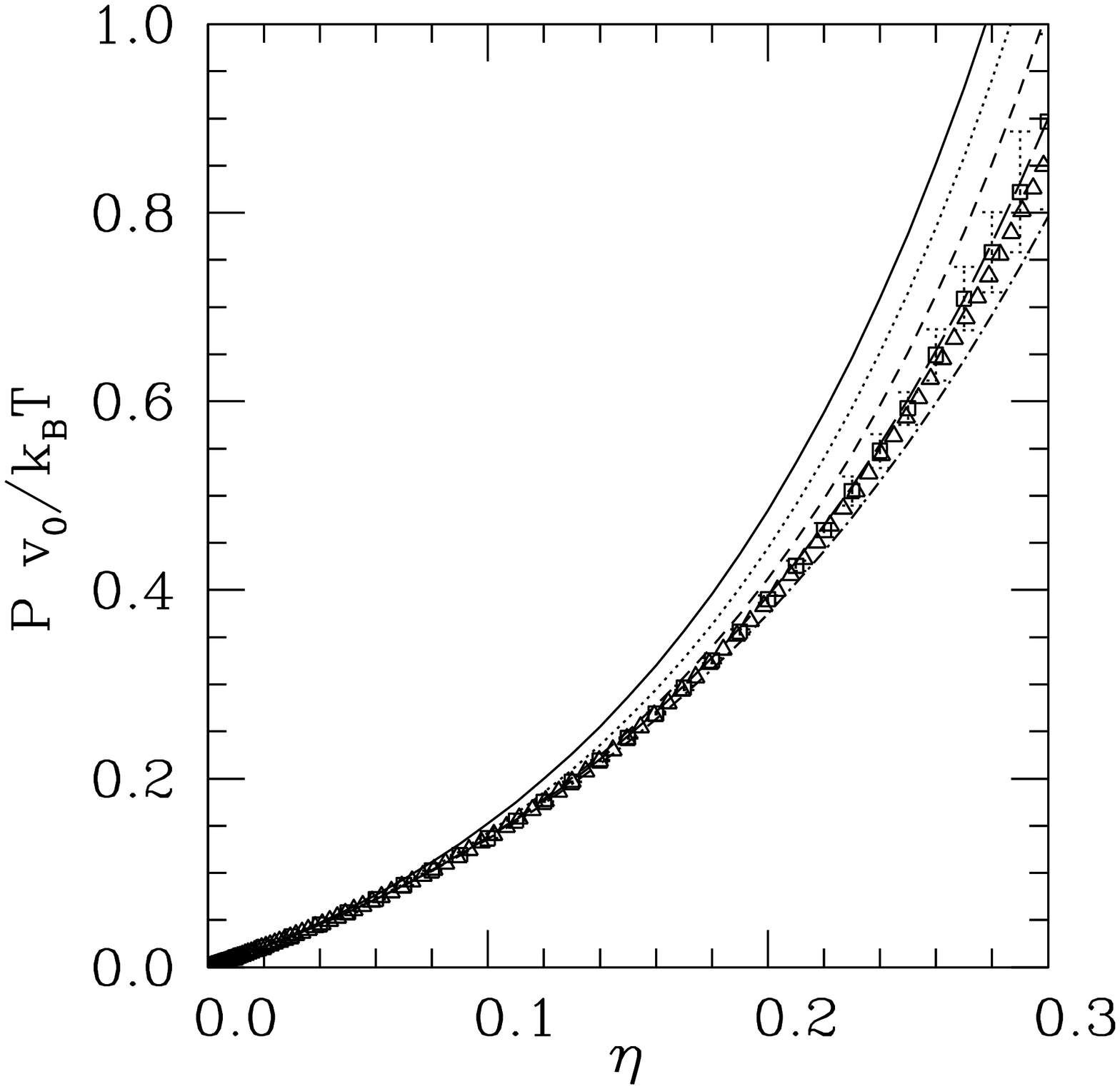}
\includegraphics[width=7cm]{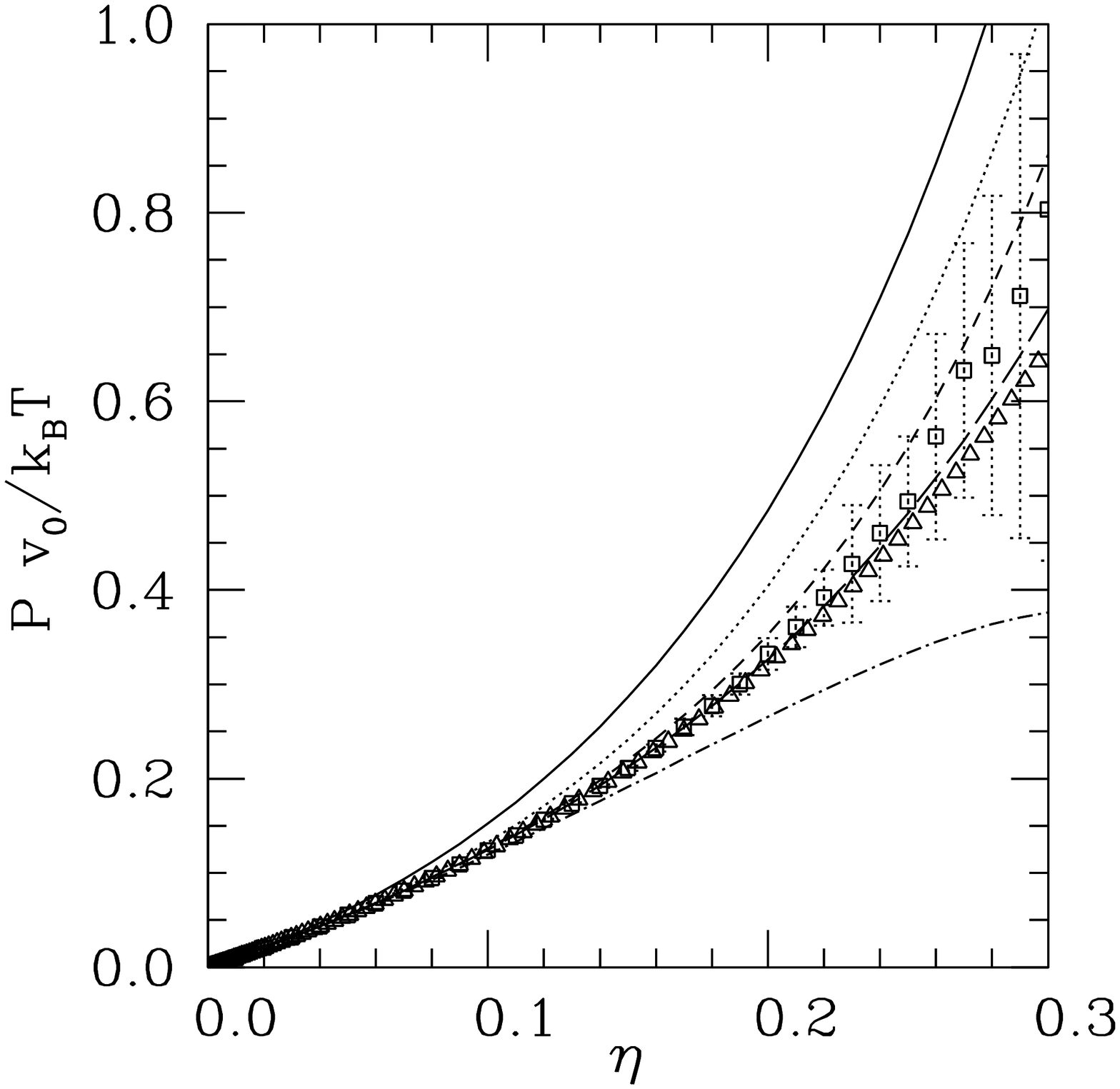}\\
\includegraphics[width=7cm]{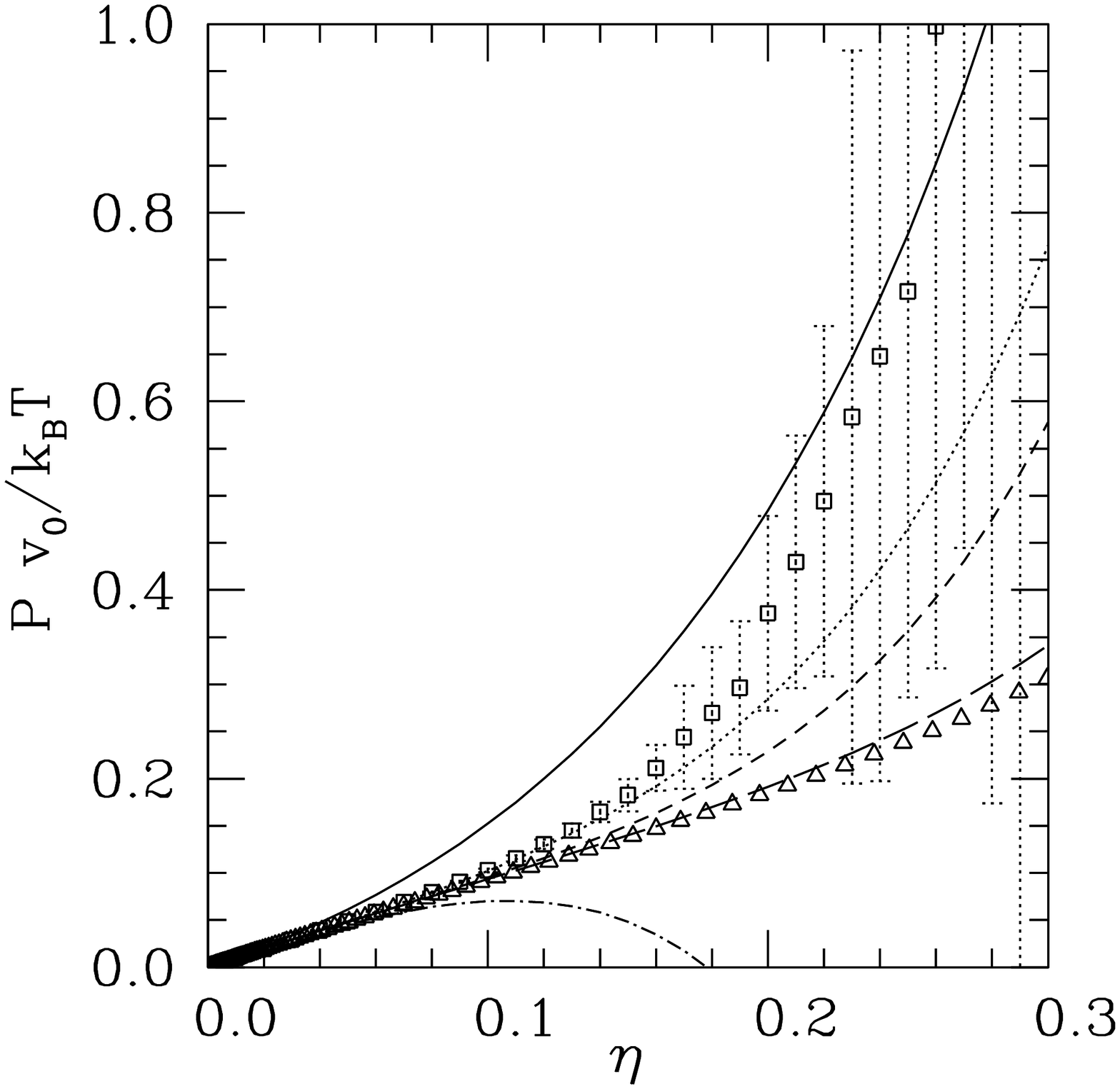}
\includegraphics[width=7cm]{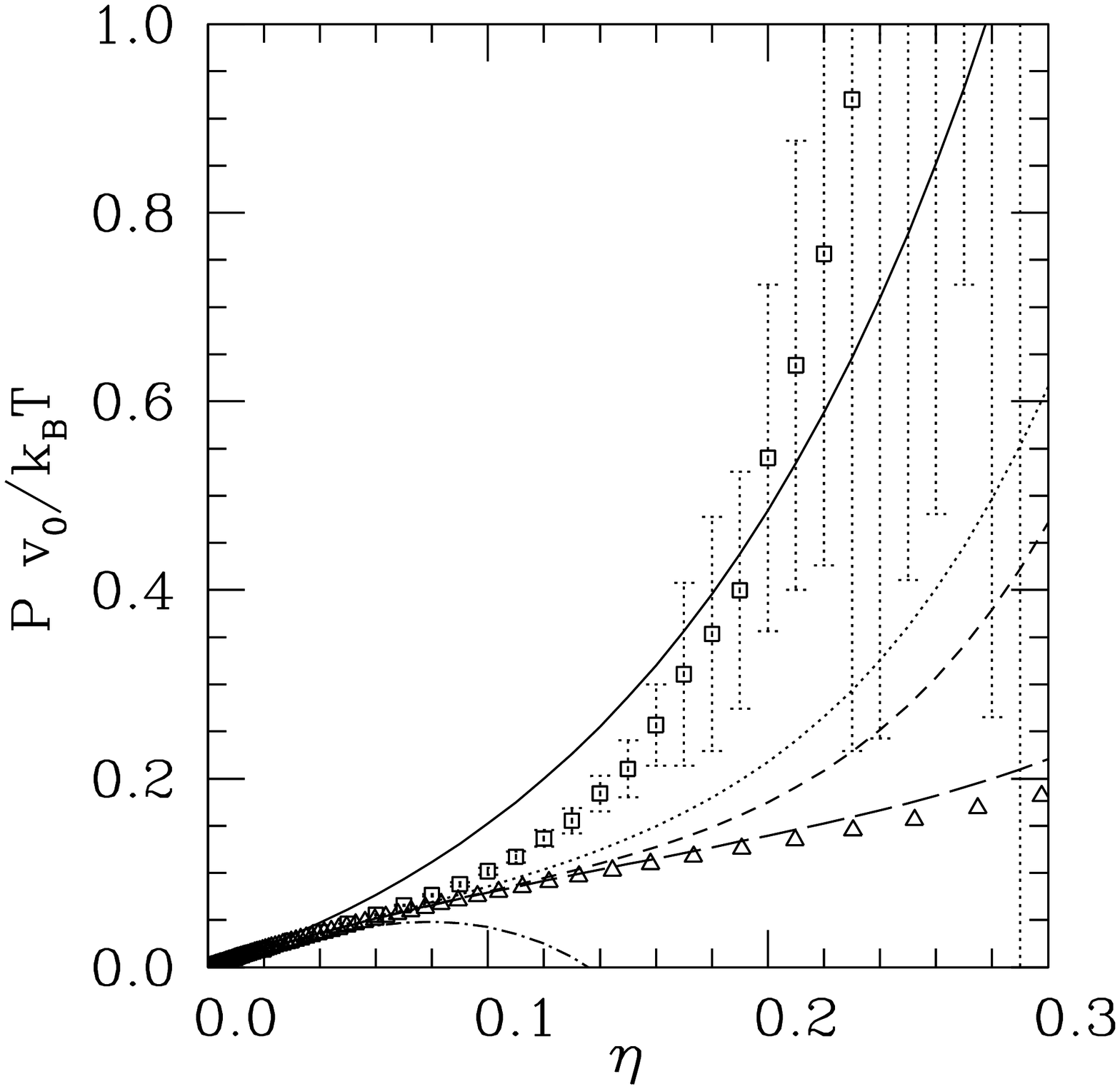}
\end{center}
\caption{}
\label{fig:oceos}
\end{figure}

We next consider the possible methods for predicting the
thermodynamic behavior of SHS fluids. As pointed out in the
introduction, a good approximation to the 
effectively exact Monte Carlo 
(MC) equation of state \cite{Miller04a} of the {\em monodisperse}
SHS model is obtained by calculating the pressure from the energy 
route within the PY approximation \cite{Watts71}. 
In the case of mixtures no comparable Monte Carlo data exists,
nor is a direct solution of the PY closure feasible, so that finding a
reliable approximation to the equation
of state remains an important open challenge.
As described in the introduction, we have tackled this in past work
within 
an approximate theory based on a density expansion of the
direct correlation function around the MSA solution
\cite{Gazzillo04,Fantoni05a,Fantoni05b}. Another possible route is
thermodynamic perturbation theory. For the Baxter SHS model it is easy
to convince oneself
that only the scheme proposed by Weeks, Chandler and Anderson (WCA)
\cite{Andersen71} 
can be applied. We have explored this possibility
in the monodisperse case, where Monte Carlo simulations provide reliable
reference data. In Fig.~\ref{fig:oceos} we compare the simulation data
with the predictions of 
the Mean-Spherical-Approximation (MSA), the modified
Mean-Spherical-Approximation (mMSA) and the C1 approximation 
(as discussed in \cite{Gazzillo04}); the results from the first and
second order WCA \cite{Andersen71}
perturbation theory are also shown. It is clear
that the mMSA and C1 approximations are fairly reliable for low and
intermediate densities, even at low reduced temperatures, while 
the second-order WCA approximation breaks down already at temperatures
significantly above the critical point ($\tau_c\approx 0.11$, depending
on the approximation used). The WCA
method therefore offers little hope of providing the 
basis for an accurate equation of state for mixtures. One also sees
readily from Fig.~\ref{fig:oceos} that the PY closure provides by far
the most accurate of all
the approximation methods. This is why we return to the problem of
solving the PY approximation for SHS mixtures in this paper.

A major challenge in calculating phase equilibria in polydisperse SHS,
or indeed any polydisperse fluid, arises from the fact that its
Helmholtz free energy is a functional 
of the distribution $p(\sigma)$ of the polydisperse
attribute~\cite{poly_review}. However, in simple systems or
approximations  this functional
dependence reduces, for the {\em
excess} free energy, to one on a finite number of moments of the
distribution. In these cases the free energy is
called truncatable~\cite{PolyPRL,AdvPhys} and the phase coexistence problem 
reduces to the solution of a finite number of coupled nonlinear
equations. For example, for the size-polydisperse SHS mixture
the mMSA and C1 approximations yield such a truncatable form for the
excess free energy involving only three moments
$\mom_1, \mom_2$ and $\mom_3$, and the two-phase coexistence
problem can easily be solved numerically \cite{Fantoni05a}. The
relevant moments are defined here including factors of density as
\be
\mom_m=\rho\int_0^\infty \sigma^m p(\sigma)\,d\sigma\ .
\label{mom_def}
\ee
for $m=1,2,3$; for later reference we note that $\mom_3$ is proportional
to the hard sphere volume fraction.

When the more accurate PY approximation is used, the presence of
polydispersity renders an analytical calculation 
of the free energy impossible (see section
\ref{sec:pyp}). In addition, even if the free energy could be
calculated in closed form, it would almost certainly
not have a truncatable form and so predictions for the phase behavior
would remain difficult to extract. We therefore propose to 
consider a small degree of polydispersity
as a perturbation \cite{Evans01} around the well-understood monodisperse
reference system (see~\cite{poly_review} for an overview of earlier
work in this perturbative spirit). Denote by $\sigma_0$ a
characteristic sphere diameter,
which will be taken as the mean diameter of the overall or ``parent''
size distribution $p^\pa(\sigma)$ in the system.
We then focus on fluids with a narrow size
distribution centered on $\sigma_0$, for which the relative particle size deviations
\be \label{deli} 
\delta=\frac{\sigma-\sigma_0}{\sigma_0}~,
\ee
are small for all particle sizes $\sigma$. Following Evans, we will
expand up to second order in these size deviations
\cite{Evans01}. The
leading order phase boundary shifts and fractionation effects then
turn out to be proportional to $s^2$, where
$s=[\langle\delta^2\rangle^\pa]^{1/2}$ 
is the normalized standard deviation -- also referred to simply as
``polydispersity'' -- of the parent distribution. Before proceeding to
the 
calculation, we address in the next section the choice of the 
stickiness coefficients $\epsilon_{ij}$ from Eq.~(\ref{tauij}). These
are irrelevant for monodisperse SHS but can have important effects on
the behavior of mixtures as we will see.

\section{The stickiness coefficients $\epsilon_{ij}$}
\label{sec:coeff}
\subsection{General arguments}

At a reduced temperature $\tau$ the Boltzmann factor
$\exp[-\phi_{ij}(r)]$ for the interaction of two SHS particles depends only on
the ratio $\epsilon_{ij}/\tau$ (and, of course, on $\sigma_{ij}$).
Physically, the
stickiness coefficients $\epsilon_{ij}$ represent
dimensionless adhesion energies between pairs of particles identified
by the species indices $i$ and $j$. (We revert to the notation for the
discrete mixture here; the same considerations obviously apply
to the polydisperse system.) The $\epsilon_{ij}$ have no analogue in the
monodisperse
case, where only the reduced temperature $\tau$ features and
$\epsilon$ can be set to unity. For (discrete or polydisperse)
mixtures, on the other hand, one needs to make an appropriate choice for the
dependence $\epsilon_{ij}={\cal F}(\sigma_i,\sigma_j)$ of the stickiness
coefficients on the particle sizes. We discuss possibilities for this
choice in this section. 

Clearly the appropriate form of the function ${\cal
F}(\sigma_i,\sigma_j)$ will depend on the kind of physical problem one
is studying. Nevertheless, it should satisfy some general requirements:
(i) Adhesion should be a purely pairwise property, and so ${\cal F}$
should depend only on $\sigma_i$ and $\sigma_j$ as anticipated by our
notation; ${\cal F}$ must clearly also be symmetric under interchange
of $\sigma_i$ and $\sigma_j$.
(ii) Since the $\epsilon_{ij}$ are dimensionless, so must ${\cal F}$
be. If it does not contain a separate lengthscale, it is therefore 
a homogeneous function of degree zero in ($\sigma_i,\sigma_j$).
The latter case is interesting because it can be seen as the sticky
limit of a scalable (i.e.\ purely size-polydisperse)
interaction~\cite{Evans01}, where by definition $\phi_{ij}(r)$ remains
unchanged when $r$, $\sigma_i$ and $\sigma_j$ are all scaled by a
common factor. (The square well potential of
Eq.~(\ref{SW}) can be put into this form by choosing
$R_{ij}=\sigma_{ij}[1+1/(A\epsilon_{ij}-1)]$; 
the sticky limit is obtained by letting $A\to\infty$.) The presence of
pure size-polydispersity has important simplifying effects on the phase
behavior~\cite{WilFasSol04,Sollich06} which we discuss further in
Sec.~\ref{sec:phases} below. 
(iii) If the adhesion depends on the surface area of the 
spheres one might expect ${\cal F}$ to depend on ratios of 
homogeneous functions of degree two in ($\sigma_i,\sigma_j$).
(iv) If the adhesive interaction vanishes when at least one of the two particles 
$i$ and $j$ degenerates to a point we need to require 
$\lim_{\sigma_i\to 0} {\cal F}(\sigma_i,\sigma_j)=0$; the limit for
$\sigma_j\to 0$ is then also zero, by the symmetry of ${\cal F}$.

In Ref.~\cite{Fantoni05a} plausibility and convenience arguments 
were adduced to suggest the following choices for the quantities
$\epsilon_{ij}$:
\begin{equation}
\epsilon _{ij}={\cal F}(\sigma_i,\sigma_j)=\left\{ 
\begin{array}{ll}
\sigma_0^{2}/\sigma _{ij}^{2} & \mbox{Case I}~, \\ 
\sigma _{i}\sigma _{j}/\sigma _{ij}^{2} & \mbox{Case II}~, \\ 
1 & \mbox{Case IV}~, \\ 
\sigma_0 /\sigma _{ij} & \mbox{Case V}~.
\end{array}
\right.   \label{Cases}
\end{equation}
Here $\sigma_0$ is a characteristic reference length for the sizes, 
taken as above to be the parental mean diameter.
In the forms originally suggested \cite{Fantoni05a}, this length was chosen
as a moment of the size distribution, $\langle \sigma^n
\rangle^{1/n}$ with either $n=1$ or 2. (Case I here corresponds to
cases I and III in Ref.~\cite{Fantoni05a}; we have kept the original
numbering for the remaining cases II, IV and V for ease of reference.)
However, this identification has the drawback of introducing many-body
effects into the pair potential, as the moments $\langle
\sigma^n\rangle$ depend upon the
thermodynamic state of the fluid, and in particular on the
concentrations of {\em all} particle species. This is why we have chosen
the fixed reference length $\sigma_0$ above, consistent with
the notion of a purely pairwise interaction.
Numerically the actual choice of $\sigma_0$ turns out to have only
very minor effects; this can be shown by calculations (not reproduced here) comparing case 
I (with fixed $\sigma_0$) with case III
from~\cite{Fantoni05a}, obtained by replacing
$\sigma_0\to \langle\sigma^2\rangle^{1/2}$.

The form of the $1/\sigma_{ij}^2$ denominator for cases I and II  in
Eq.~(\ref{Cases}) is forced by technical constraints detailed
in Ref.~\cite{Fantoni05a}, but these still leave some flexibility in the
choice of numerator; cases I and II assume respectively a mean-field-like
and a decoupled 
dependence between stickiness and size. Case IV corresponds to the choice of
constant coefficients (independent 
of particle sizes), while case V was selected in
Ref. \cite{Fantoni05a} specifically to permit analytical solution
within the C1 closure. Note that not all four cases have all of the properties
(ii--iv) listed above as possible requirements. E.g.\ only cases II and IV
are homogeneous functions of ($\sigma_i,\sigma_j$) as required by (ii)
when no additional lengthscale such as $\sigma_0$ is involved; they
are therefore purely size-polydisperse. The
properties (iii) and (iv) hold only for case
II. It can be argued \cite{Gazzillo06} that the dependence on
$\sigma_i\sigma_j/\sigma_{ij}^2$ assumed in case II is quite 
generic for solutions of colloids, micelles or globular proteins,
at least in the high-temperature regime where a
linearized approximation for the Boltzmann factor is sufficient. While
this favors case II, for phase coexistence we are interested in lower
temperatures where it is less clear which case is physically more
appropriate; we will therefore include all four cases in our analysis.

For our perturbative analysis we only need to know the coefficients in
the expansion of the $\epsilon_{ij}$ around the typical
particle size $\sigma_i=\sigma_j=\sigma_0$, up to quadratic order in
the relative particle size deviations $\delta_i=(\sigma_i-\sigma_0)/\sigma_0$:
\be
\epsilon_{ij}=\epsilon_0+\epsilon_{1a}(\delta_i+\delta_j)+
\epsilon_{2a}\delta_i\delta_j+\epsilon_{2b}(\delta_i^2+\delta_j^2)+\ldots
\label{epsij_expansion}
\ee
The coefficients $\epsilon_0$, $\epsilon_{1a}$, $\epsilon_{2a}$,
$\epsilon_{2b}$ of this expansion are given in
Table~\ref{tab:peps} for the four cases listed above. Note that
$\epsilon_0=1$ always so that in the monodisperse limit the
$\epsilon_{ij}$ are irrelevant as they should be.

\subsection{Stickiness coefficients for the Asakura-Oosawa model}
\label{sec:AO}

So far we have considered choices for the stickiness coefficients suggested 
by rather general arguments. One may wonder whether the
$\epsilon_{ij}$ can be derived more directly from a physical picture. 
We shall pursue this here for the well-known
Asakura-Oosawa model of colloid-polymer mixtures, which for small
polymers leads to a short-ranged attractive depletion potential acting
between the colloids \cite{Asakura54}. We shall show that, while a formal
sticky limit cannot be taken in general when colloids of different
sizes are present, an effective SHS model
can still be derived when the polymer size is small but kept
nonzero. This then simplifies further in the perturbative
approach for weak polydispersity adopted here.

\begin{figure}[ht!]
\begin{center}
\includegraphics[width=11cm]{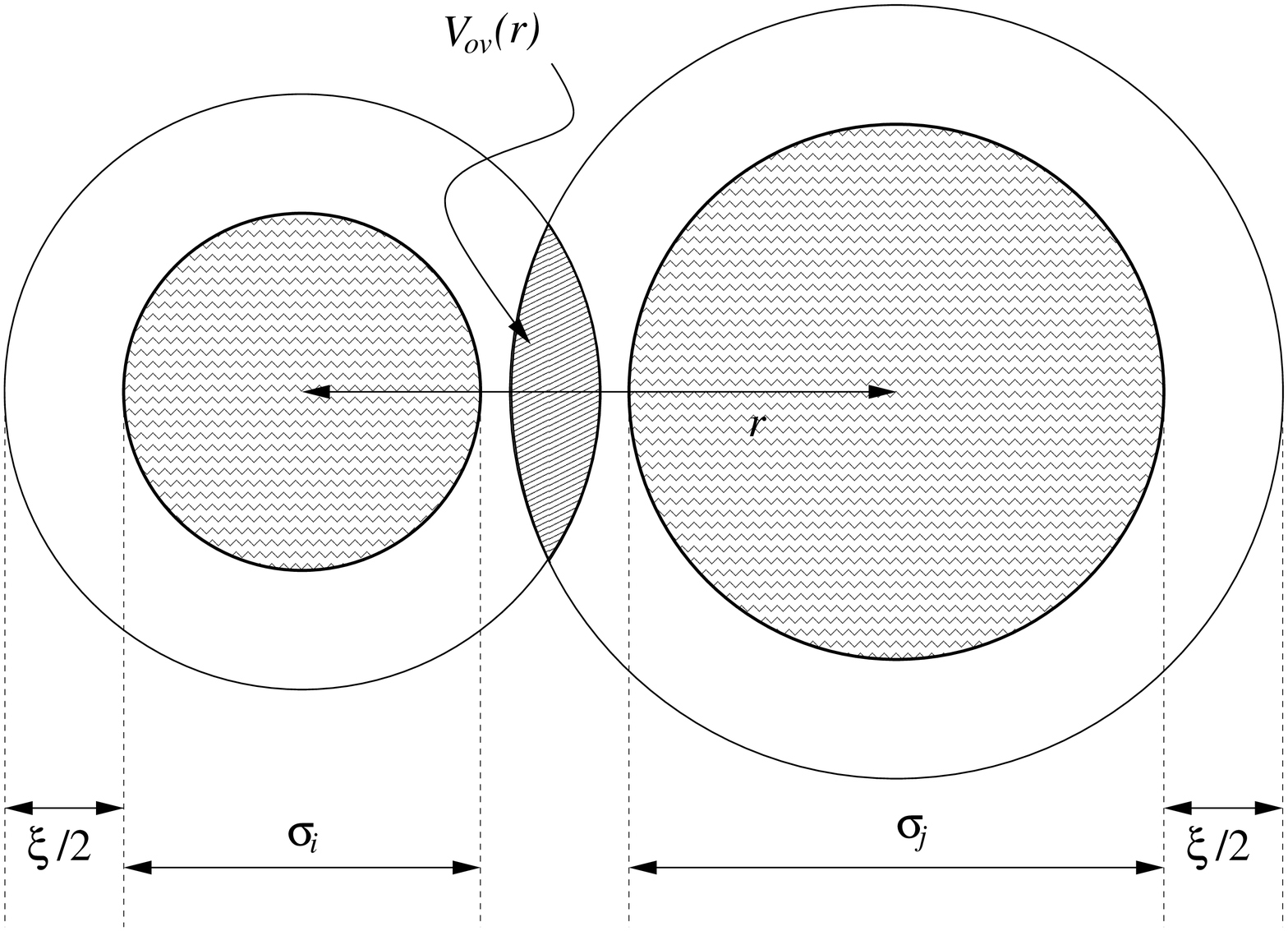}
\end{center}
\caption{}
\label{fig:ao}
\end{figure}

Consider two colloidal particles represented by 
impenetrable spheres of diameter $\sigma_i$ and $\sigma_j$ 
immersed in a solution of non-interacting polymers. Within the
Asakura-Oosawa model, the polymers are simplified to spheres of
diameter $\xi$ which can fully penetrate each other 
but have a hard sphere interaction with the colloids.
It is well known that such a system develops an entropically driven effective attraction 
between the colloidal particles. This arises due to a reduction in the
volume from which the polymers are excluded when the exclusion
zones around the colloids overlap (see Fig.~\ref{fig:ao}). This
overlap volume as a function of the distance $r$ between the sphere
centers is
\bq
{\cal V}_{\rm ov}(r)=\frac{\pi}{12}\left[r^3-6(R_i^2+R_j^2)r+
8(R_i^3+R_j^3)-3(R_i^2-R_j^2)^2\frac{1}{r}\right]
\theta(\sigma_{ij}+\xi-r)
\eq
where $R_k=(\sigma_k+\xi)/2$ and only distances $r>\sigma_{ij}$ are
allowed because of the hard colloid-colloid repulsion. The effective
colloid-colloid attraction induced by the presence of the polymers is
then just the overlap volume times the polymer osmotic pressure
\cite{Asakura54,Hansen-Barrat}, giving the overall AO interaction potential
\bq
\pot^{\rm AO}_{ij}(r)=\left\{ 
\begin{array}{ll}
+\infty & 0<r<\sigma _{ij}~, \\ 
-\rho_p {\cal V}_{\rm ov}(r) & \sigma_{ij}\leq r < \sigma_{ij}+\xi~, \\ 
0 & r\geq \sigma_{ij}+\xi~,
\end{array}
\right.
\eq
This expression can be obtained formally by integrating out the
polymer degrees of freedom from the partition function at fixed
polymer chemical potential. The latter is conveniently parametrized by
the density $\rho_p$ of polymers in a reservoir connected to the
system; because the polymers are taken as ideal, their osmotic
pressure is then $\kT\rho_p$ and the $\kT$ is absorbed by our choice
of units. The effective colloid-colloid
interaction will in general contain also many-body terms, but these
vanish in the limit of small polymers (for monodisperse colloids the
condition is $\xi<0.1547\sigma_0$) that we are interested in.

To map to an equivalent SHS potential, which should be physically
reasonable for small polymer-to-colloid size ratio $\xi/\sigma_0$, one
equates the corresponding second virial coefficients. The hard core
makes the same contribution $(B_{2,{\rm HS}}^{ij}=2\pi\sigma_{ij}^3/3)$ in
the SHS and the original AO potential, so one can focus on the
normalized deviation of the second virial coefficient from this HS value,
\bq \nonumber
\Delta B_{2,{\rm AO}}^{ij}&=&
\frac{B_{2,{\rm AO}}^{ij}-B_{2,{\rm HS}}^{ij}}{B_{2,{\rm HS}}^{ij}}\\
\nonumber
&=&\frac{3}{\sigma_{ij}^3}\int_{\sigma_{ij}}^{\sigma_{ij}+\xi}
\left[1-e^{-\pot^{\rm AO}_{ij}(r)}\right]r^2\;dr~.
\eq
For the SHS potential this quantity equals $-1/(4\tau_{ij})$, so the
stickiness parameters in the mapped SHS system are assigned as
\bq
 \label{db2}
\frac{1}{12\tau_{ij}} =
\frac{1}{\sigma_{ij}^3}\int_{\sigma_{ij}}^{\sigma_{ij}+\xi} 
\left[e^{-\pot^{\rm AO}_{ij}(r)}-1\right]r^2\;dr~.
\eq
We now proceed to simplify this expression for small $\xi$; in the
limit $\xi\to0$, the original AO model should become fully
equivalent to the mapped SHS system. We will see that for mixtures of
colloids of different sizes this strict mathematical limit cannot be
taken consistently; nevertheless, as long as $\xi/\sigma_0$ is small,
we expect the SHS mixture to give a reasonably accurate description
of the underlying AO model.

To simplify Eq.~(\ref{db2}) we change
integration variable from $r$ to $z=(r-\sigma_{ij})/\xi$,
expand the attractive tail of the AO
potential in $\xi$ as
\bq
-\pot^{\rm AO}_{ij}(z)=\frac{\pi}{4}\rho_p\xi^2
\frac{\sigma_i\sigma_j}{\sigma_{ij}}(z-1)^2+O(\xi^3)~.
\eq
and retain only the leading term. Similarly approximating
$r^2=(\sigma_{ij}+\xi z)^2=\sigma_{ij}^2+O(\xi)$ yields
\bq
\frac{1}{12\tau_{ij}} =
\frac{\xi}{\sigma_{ij}} \int_0^1 \left[e^{\gamma_{ij}(1-z)^2}-1\right]dz
= \frac{\xi}{\sigma_{ij}}
\left[\frac{1}{2}\sqrt{\frac{\pi}{\gamma_{ij}}}
\mbox{erfi}(\sqrt{\gamma_{ij}})-1\right]
\label{db2b}
\eq
where
\[
\gamma_{ij}=\frac{\pi}{4}\rho_p\xi^2\frac{\sigma_i\sigma_j}{\sigma_{ij}}
\]
is the value of the attractive potential at contact and
$\mbox{erfi}(z)=\mbox{erf}(iz)/i$ is the imaginary
error function.  Because of the prefactor $\xi/\sigma_{ij}$ in
Eq.~(\ref{db2b}), $\gamma_{ij}$ has to grow as $\xi$ decreases if we
want to keep
$\tau_{ij}$ finite. For large argument the error function behaves as
$\mbox{erfi}(z)=e^{z^2}[1/z+O(1/z^3)]/\sqrt{\pi}$ and so
\bqn
\frac{1}{12\tau_{ij}} \approx
\frac{\xi}{\sigma_{ij}}\frac{e^{\gamma_{ij}}}{2\gamma_{ij}}
=\frac{2}{\pi\rho_p\xi\sigma_i\sigma_j}
e^{\frac{\pi}{4}\rho_p\xi^2 \frac{\sigma_i\sigma_j}{\sigma_{ij}}}~.
\eqn
A nonzero limit value of $\tau_{ij}$ for $\xi\to 0$ thus requires that
$\gamma_{ij}$ grows logarithmically as
$\gamma_{ij}=\ln(\sigma_{ij}/\xi)$ to leading order. The corresponding
polymer reservoir density, likewise to leading order, goes as 
\bq \label{rhop}
\rho_p=\frac{4}{\pi}\frac{\sigma_{ij}}{\sigma_{i}\sigma_{j}}
\frac{\ln(\sigma_0/\xi)}{\xi^2}~.
\eq
The dominant dependence $\rho_p\propto \xi^{-2}$ in this expression
arises because the value of the AO potential at contact scales as
$\rho_p\xi^2$; the additional logarithmic factor increases this
interaction strength to compensate for the decreasing range of the
attraction as $\xi\to 0$. Note that even though the polymer density
diverges, the polymers 
do in fact become very dilute as one sees from the (reservoir) volume
fraction 
$\eta_p=(\pi/6)\rho_p\xi^3\sim \xi\ln(\sigma_0/\xi)$ occupied by the polymer
spheres.

For monodisperse colloids, the above procedure produces an unambiguous
sticky limit for $\xi\to 0$. The explicit form of Eq.~(\ref{rhop})
shows, however, that this limit cannot be taken straightforwardly for
mixtures: the prefactors $\sigma_{ij}/(\sigma_i \sigma_j)$ of the
required leading order divergences of the polymer density are incompatible
with each other for different pairs of particle species. In other
words, if the $\xi$-dependence of the polymer density is chosen to
keep one specific $\tau_{ij}$ finite and nonzero, then the others
would either tend to zero or grow to infinity in the sticky limit. The
example of a binary mixture illustrates this. Suppose that
$\sigma_1>\sigma_2$ and that the polymer density is tuned to keep the
$\tau_{11}$ finite. Then $1/\tau_{12}$ and $1/\tau_{22}$ would both
tend to zero for $\xi\to 0$ so that all interactions involving
particles of species 2 become purely HS-like, without any attractive
contributions (this is system B studied in \cite{Fantoni05b}).

In the absence of a strict sticky limit, we will content ourselves
with applying the mapping (\ref{db2b}) for small but nonzero
polymer-to-colloid size ratios $\xi/\sigma_0$. The properties of the
resulting SHS mixture should then still give a good approximation to
those of the original AO model. In the perturbative setting 
of this paper we can then expand Eq.~(\ref{db2b}) in the small relative
deviations $\delta_i=(\sigma_i-\sigma_0)/\sigma_0$ of the particle
sizes from the parental mean. In the decomposition
$1/\tau_{ij}=\epsilon_{ij}/\tau$ of Eq.~(\ref{tauij}) we fix the scale
of the $\epsilon_{ij}$ by requiring as before that $\epsilon_{ij}=1$ for particles
of the reference size $\sigma_i=\sigma_j=\sigma_0$. This gives
\be
\frac{1}{\tau} = \frac{12\xi}{\sigma_0}
\left[\frac{1}{2}\sqrt{\frac{\pi}{\gamma}}
\mbox{erfi}(\sqrt{\gamma})-1\right] \approx 
\frac{6\xi}{\sigma_0}\frac{e^{\gamma}}{\gamma}
\label{tauAO}
\ee
for the reduced temperature,
where 
\bqn
\gamma=\frac{\pi}{4}\rho_p\xi^2\sigma_0~.
\eqn
The second, approximate equality in Eq.~(\ref{tauAO}) holds for large
$\gamma$ as before. To find the perturbative expansion of the
stickiness coefficients $\epsilon_{ij}$, we note first that the
potentials at contact expand as 
\bqn
\gamma_{ij}=\gamma\left[1+\frac{1}{2}(\delta_i+\delta_j)
+ \frac{1}{2}\delta_i\delta_j-\frac{1}{4}(\delta_i^2+\delta_j^2)\right]
\eqn
Since the erfi in Eq.~(\ref{db2b}) grows at most as
$\exp(\gamma_{ij})$, a second order Taylor expansion will give an
accurate approximation as long as the perturbations in $\gamma_{ij}$
are $\ll 1$. This requires $\delta_i\ll 1/\gamma$, which then automatically 
enforces $\delta_i\ll 1$ since we expect $\gamma$ to be at least of
order unity for the mapping to a SHS mixture to make sense. Under
these conditions one then has a valid perturbation expansion of the
$\epsilon_{ij}$. The coefficients defined in
Eq.~(\ref{epsij_expansion}) are found as $\epsilon_0=1$ (by our choice
of $\tau$) and 
\[
\epsilon_1=\frac{-1+g_1}{2}, \quad
\epsilon_{2a} = \frac{1+g_2}{2}, \quad
\epsilon_{2b} = \frac{1-2g_1+g_2}{4}
\]
where
\bqn
g_1 &=& \frac{e^\gamma-1}{\sqrt{\pi/\gamma}\,\mbox{erfi}(\sg)-2} -
\frac{1}{2}
\\
g_2 &=& \frac{[3+e^\gamma(2\gamma-3)]/4}{\sqrt{\pi/\gamma}\,\mbox{erfi}(\sg)-2} +
\frac{3}{8}~.
\eqn
From Eq.~(\ref{tauAO}) one sees that the reduced temperature is set
by the contact potential $\gamma$, which itself is proportional to the polymer
reservoir density. Unlike the more ad-hoc choices of
Eq.~(\ref{Cases}), the
expansion of the $\epsilon_{ij}$ in terms of the $\delta_i$ depends on
the reduced 
temperature $\tau$, via $\gamma$. For large $\gamma$ one can use the
leading order approximations $g_1\approx\gamma - 1$, $g_2\approx
(\gamma^2-2\gamma+1)/2$ to evaluate this dependence. However, since
typical values of $\gamma$ are only 
logarithmically large in $\sigma_0/\xi$ it is generally safer to work
with the full expressions.

\section{Perturbation theory for the polydisperse PY closure}
\label{sec:core}

In this section we come to the core of our analysis. We first review Evans'
perturbative framework for slightly polydisperse systems. To apply
this to the PY approximation for SHS mixtures we will need 
the perturbative expansion of certain correlation function values at
contact; from these we can then finally find the excess free energy.

\subsection{Evans' perturbative expansion}
\label{sec:evans}

The starting point for an analysis of the phase behavior of
polydisperse systems is the excess free energy density. In
general this is a functional of the size distribution $p(\sigma)$ in
the system. It is also a function of the particle density $\rho$, and
of temperature; we do not write the latter explicitly below. For slightly
polydisperse systems it is expedient to switch from $\sigma$ to the
relative deviations $\delta$ from the reference size $\sigma_0$. By
the fundamental assumption of a narrow size distribution, the $\delta$
are small quantities, and one can expand the excess free energy
density $f\ex$, measured again in units of $\kT$, in terms of moments of
$p(\delta)$ \cite{Evans01}:
\be
f\ex(\rho,[p(\delta)])=
f\ex\mono(\rho)+\rho a(\rho)\avdel+\rho b(\rho)\avdels+
\rho c(\rho)\avdel^2+\ldots ~.
\label{evans-free}
\ee
Here terms up to second order in $\delta$ have been retained; these
give the leading effects on the phase boundaries \cite{Evans01}. Our
functions $a,b,c$ differ by factors of $\rho$ from those defined in
Ref.~\cite{Evans01}, so that e.g.\ $a$ equals Evans' $A/\rho$; this
simplifies the statement of Eqs.~(\ref{evans-shift}-\ref{evans-frac})
below. The leading term $f\ex\mono$ is 
the excess free energy density of the monodisperse
reference system where all particles have $\delta=0$.

Given the above expansion of the excess free energy, the conditions
for two-phase equilibria of the near-monodisperse fluid can be solved
perturbatively \cite{Evans01}. We briefly recall the main results.
The fluid is initially
in a parent phase of density $\rho^\pa$, with a parent size distribution
function $p^\pa(\delta)$, where $\langle\delta\rangle^\pa=0$ by our
choice of the reference size $\sigma_0$ as the parental mean.
In order to lower its free
energy, the fluid can split into 
two daughter phases of densities $\rho\one$ and $\rho\two$, with
distribution functions $p\one(\delta)$ and $p\two(\delta)$ which are
in general different from the parent distribution, a phenomenon referred to as
{\sl fractionation}~\cite{poly_review}. The densities and size
distributions can be worked out perturbatively at any point inside the
coexistence region \cite{Evans01}; we focus on the properties at the
{\em onset} of phase coexistence, which are most easily accessible
experimentally.

Suppose the system is just starting to phase separate, with all of the
volume except for an infinitesimal fraction still occupied by phase 1,
with density $\rho\one$. Conservation of particle number then requires
that $p\one(\delta)=p^\pa(\delta)$, i.e.\ the size distribution in
this {\em cloud} phase equals the parent. The coexisting {\em
shadow} phase 2, on the other hand, will generally have
$p\two(\delta)\neq p^\pa(\delta)$. Evans \cite{Evans01} showed that
the cloud and shadow densities,
$\rho\one=\rho\one\mono+\delta\rho\one$ and
$\rho\two=\rho\two\mono+\delta\rho\two$, are shifted from their
monodisperse values $\rho\one\mono$ and $\rho\two\mono$ by
\bq
\delta\rho\one &=& -s^2\rho\one\mono\kappa(\rho\one\mono)\left[(\rho\one\mono)^2 b'(\rho\one\mono) +
  \frac{(\Delta a)^2 + 2\Delta b}{2\Delta(1/\rho)}\right]
\label{evans-shift}
\\
\delta\rho\two &=& -s^2\rho\two\mono\kappa(\rho\two\mono)\left[(\rho\two\mono)^2 b'(\rho\two\mono) +
  \frac{(\Delta a)^2 + 2\Delta b}{2\Delta(1/\rho)} +
  (\rho\two\mono)^2 a'(\rho\two\mono) \Delta a\right]
\label{evans-shadow}
\eq
Here $a'\equiv\partial a/\partial\rho$, $b'\equiv\partial b/\partial\rho$ and
$\kappa(\rho)=1/[\rho+\rho^2 (\partial/\partial\rho)^2 f\ex\mono(\rho)]$ is
the isothermal 
compressibility of the monodisperse reference system.
The shorthand $\Delta$ indicates differences between the two
monodisperse reference phases, e.g.\ $\Delta a =
a(\rho\one\mono)-a(\rho\two\mono)$. Finally, recall that $s$ is the
parent polydispersity: the phase boundary shifts are to leading order
quadratic in $s$.

It is worth noting that Eqs.~(\ref{evans-shift},\ref{evans-shadow}) are not
symmetric in $\rho\one\mono$ and $\rho\two\mono$; by interchanging the two
densities one therefore obtains a different cloud-shadow
pair. Physically, this corresponds to approaching the onset of phase
separation from low or high densities; in a polydisperse system the
coexisting phases are different in the two situations since only the
respective majority (cloud) phase has the parental size distribution.
The size distribution in the corresponding {\em shadow} reads, to leading
order in $\delta$ \cite{Evans01},
\be
\label{evans-frac}
p\two(\delta)=p^\pa(\delta)[1+(\Delta a)\delta]~.
\ee
Overall, the monodisperse binodal delimiting the coexistence region
splits into separate cloud and shadow curves, which intersect in the
critical point \cite{poly_review}.
Quantitative information about the critical region is not accessible within the
perturbative expansion of
Eqs.~(\ref{evans-shift},\ref{evans-shadow}), however, since the
compressibility $\kappa$ diverges as the critical point is approached.

The above summary shows that knowledge of the functions $a$, $b$ and $c$
is sufficient to calculate the leading order phase boundary shifts and
fractionation effects for weakly polydisperse systems. In the next two
subsections we calculate these functions for the SHS mixture within
the PY approximation.

\subsection{Perturbative analysis of the PY closure}
\label{sec:py}
\label{sec:pyp} 

To lighten the notation in the rest of the paper, we make all
densities dimensionless by measuring them in units of $v_0^{-1}$,
where
\[
v_0=(\pi/6)\sigma_0^3
\]
is the volume of a particle with the
reference diameter. The third moment $\mom_3$ defined in
Eq.~(\ref{mom_def}) is then identical to the hard sphere volume
fraction $\eta$. 
We also measure all particle sizes $\sigma$ in terms of
$\sigma_0$, so that the relation between $\sigma$ and the fractional
deviation from the parental mean diameter becomes simply
$\sigma=1+\delta$. In the monodisperse case, where all particles have
$\delta=0$, all moments (\ref{mom_def}) are then identical and equal to the
density $\rho$ (which also equals the volume fraction $\eta$).
Finally, for notational simplicity we again revert temporarily to the case of a
discrete $p$-component 
SHS mixture; the final results will be expressed in terms of averages
over the size distribution and so generalize immediately to
fully polydisperse systems.

In order to extract the desired thermodynamic quantities
from the PY closure, the following set of $p(p+1)/2$ coupled
quadratic equations needs to be solved first \cite{Perram75}, 
\bq \label{pss}
L_{ij}=\alpha_{ij}+\beta_{ij}\sum_mx_m\left[\frac{1}{12}L_{im}L_{jm}-
\frac{1}{2}\left(L_{im}\phi_{mj}+L_{jm}\phi_{mi}\right)\right]~,
~~~i,j=1,2,\ldots,p
\eq
where the unknowns are
\bqn
L_{ij}=\frac{y_{ij}(\sigma_{ij})\sigma_{ij}^2\epsilon_{ij}}{\tau}~.
\eqn
Here $y_{ij}(\sigma_{ij})$ is the partial cavity function at contact
which is proportional to the probability of finding a particle of
species $j$ touching any given particle of species $i$. In
Eq.~(\ref{pss}) the coefficients $\alpha_{ij}$, $\beta_{ij}$, and
$\phi_{ij}$ are given by 
\bq \label{ab}
\alpha_{ij}&=&y^{\rm HS}_{ij}(\sigma_{ij})\sigma_{ij}^2\epsilon_{ij}/\tau~,\\
\beta_{ij}&=&\rho\sigma_{ij}\epsilon_{ij}/\tau~,\\ \label{phi}
\phi_{ij}&=&\sigma_i\sigma_j/\Delta~.
\eq
Here the quantities
\bq \label{yHS}
y^{\rm HS}_{ij}(\sigma_{ij})=\frac{1}{\Delta}+
\frac{3}{2}\frac{\mom_2}{\Delta^2}
\frac{\sigma_i\sigma_j}{\sigma_{ij}}
\eq
are the PY partial cavity functions at contact for the HS fluid (to which
the SHS fluid reduces at infinite reduced temperature $\tau$) 
and we abbreviate $\Delta=1-\eta$, with $\eta\equiv\mom_3$ the HS packing
fraction as before.
Notice that all four sets of coefficients $L_{ij}, \alpha_{ij}, \beta_{ij}$,
and $\phi_{ij}$ are symmetric under exchange of the species indices
$i$ and $j$.

For one-component fluids, the system (\ref{pss}) reduces to a single
quadratic equation. Baxter \cite{Baxter68} showed that only the
smaller of the 
two real solutions (provided such solutions exist at all) is physically
significant; it is given explicitly in Eq.~(\ref{L0}) below.
For true mixtures ($p>1$), explicit solution of the rather complicated system (\ref{pss})
of algebraic equations is feasible at best numerically   
(except for special cases \cite{Robertus89,Fantoni05b}) and is the
computational bottleneck of the PY solution. For large $p$, and
certainly for the polydisperse limit $p\to\infty$, it is
impossible in practice. However, progress can be made for near-monodisperse fluids
by solving~(\ref{pss}) perturbatively. The $L_{ij}$ will generically
depend on the reduced temperature $\tau$, the overall number
density $\rho$, the sizes $\sigma_i$ and $\sigma_j$ of the particles
at contact, and all the molar fractions $x_i$ (or their polydisperse
analogue, the size distribution $p(\delta)$). For small $\delta_i$ we
can therefore expand to quadratic order as
\bq \nonumber
L_{ij}&=&L_0+L_{1a}(\delta_i+\delta_j)+
L_{1b}\avdel+\\ \label{peps} 
&&L_{2a}\delta_i\delta_j+L_{2b}(\delta_i^2+\delta_j^2)
+L_{2c}\avdel(\delta_i+\delta_j)+L_{2d}\avdel^2
+L_{2e}\avdels~.
\eq
The idea now is to insert this expansion, and the analogous expansions
of the known coefficients $\alpha_{ij}$, $\beta_{ij}$ and $\phi_{ij}$,
into the r.h.s.\ of Eq.~(\ref{pss}). Having done this, one re-expands to
quadratic order in $\delta_i$, $\delta_j$, $\delta_m$ and
$\langle\delta\rangle$, and to linear order in
$\langle\delta^2\rangle$. Finally one replaces $\sum_m x_m=1$ and
$\sum_m x_m \delta_m^n=\langle\delta^n\rangle$ for $n=1,2$. Comparing
terms of the same form on the left and right of Eq.~(\ref{pss}) one
then finds a relatively simple set of equations for the coefficients
$L_0$, \ldots, $L_{2e}$, as outlined in
the Appendix. To order zero in polydispersity one of course retrieves
Baxter's original quadratic equation (Eq.~(\ref{o0})), whose
physically relevant solution is
\bq
L_0=\frac{\alpha_0}{\displaystyle\frac{1}{2}
\left[1+\frac{\beta_0}{\Delta_0}
+\sqrt{\left(1+\frac{\beta_0}{\Delta_0}\right)^2-
\frac{\beta_0\alpha_0}{3}}\right]}~,
\label{L0}
\eq
where $\Delta_0=1-\rho$ is the value of $\Delta$ in a monodisperse
system with density $\rho$. Since we are perturbing around
the physical 
solution~(\ref{L0}) for the monodisperse case, the results we find for slightly
polydisperse 
mixtures will automatically have the correct physical behavior. In a
non-perturbative solution, one would need to check separately that the
solution branch with the correct low-density limit
$L_{ij}\to{\sigma_{ij}^2}/{\tau_{ij}}$ has been selected; this
condition arises since $y_{ij}(\sigma_{ij})\to 1$ at low density. 

The conditions imposed by Eq.~(\ref{pss}) for the higher order
expansion coefficients $L_{1a}$, \ldots, 
$L_{2e}$ turn out to be {\em linear} and can be straightforwardly
solved order by order; see the Appendix. The region in the
density-temperature plane where Eq.~(\ref{pss}) has no physical solution
therefore remains as in the monodisperse case, being delimited by
$\rho_-<\rho<\rho_+$ with 
\be
\rho_{\pm}=\frac{1-6(\tau-{\tau}^2)\pm
\sqrt{1-12\tau+18{\tau}^2}}{5-12\tau+6{\tau}^2}~.
\label{existence}
\ee
for $\tau<(2-\sqrt{2})/6$. This is clearly an artifact of our
finite-order perturbation theory, given that we
know from numerical solutions of Eq.~(\ref{pss}) that the 
region where solutions exist does change with increasing polydispersity 
\cite{Robertus89}. To reproduce this effect within our approach, a
resummation of the 
perturbation theory to all orders would be needed.

\subsection{Excess free energy}
\label{sec:thermodynamics}

Given the perturbative expansion for $L_{ij}$, we can determine the
free energy of weakly polydisperse SHS mixtures in the PY approximation.
There are three known thermodynamic routes (via the energy, 
compressibility, and virial) that could potentially be used \cite{Barboy79}. 
We focus on the one that gives the most
reliable equation of state for the monodisperse system (see
Fig.~\ref{fig:oceos}), i.e.\ the energy route.
It predicts in general for the $\tau$-derivative of the excess free energy
density
\[ 
\frac{\partial f\ex}{\partial\tau}=\frac{\rho^2}{\tau}
\sum_{ij}x_ix_j\sigma_{ij}L_{ij}
\]
Inserting the expansion (\ref{peps}) of $L_{ij}$ and re-expanding to
quadratic order yields
\[
\frac{\partial f\ex}{\partial\tau}=\frac{\rho^2}{\tau}
\left[\Gamma_0+\Gamma_1\avdel+\Gamma_2\avdel^2
+\Gamma_3\avdels\right]~,
\]
where
\bqn
\Gamma_0&=&L_0~,\\
\Gamma_1&=&L_0+2L_{1a}+L_{1b}~,\\
\Gamma_2&=&L_{1a}+L_{1b}+L_{2a}+2L_{2c}+L_{2d}~,\\
\Gamma_3&=&L_{1a}+2L_{2b}+L_{2e}~.
\eqn
We can then integrate from the desired value of $\tau$ to the
hard-sphere limit $\tau\to\infty$ to find
\bqn
\Delta {f}\ex\equiv {f}\ex-{f}\ex_{\rm HS}=
\Delta {f}\ex_0+
\Delta {f}\ex_1\avdel+\Delta {f}\ex_2\avdel^2+
\Delta {f}\ex_3\avdels~,
\eqn
where 
\bqn
\Delta {f}\ex_i= - \rho^2 \int_\tau^\infty
\Gamma_i(\tau^\prime)\frac{d\tau^\prime}{\tau^\prime}~,
~~~i=0,1,2,3~.
\eqn
and ${f}\ex_{\rm HS}$ is the excess free energy density of the HS
fluid. For the latter we use the standard Boublik 
\cite{Boublik70}, Mansoori, Carnahan, Starling, and Leland 
\cite{Mansoori71} (BMCSL) expression \cite{note2}. Expanded to
second order in polydispersity this reads
\bqn
{f}\ex_{\rm HS}={f}\ex_{{\rm HS},0}+
{f}\ex_{{\rm HS},1}\avdel+
{f}\ex_{{\rm HS},2}\avdel^2+
{f}\ex_{{\rm HS},3}\avdels~,
\eqn
where
\bqn
{f}\ex_{{\rm HS},0}&=&\displaystyle\frac{\rho^2(4-3\rho)}
{\Delta_0^2}~,\\ 
{f}\ex_{{\rm HS},1}&=&\displaystyle\frac{6\rho^2(2-\rho)}
{\Delta_0^3}~,\\
{f}\ex_{{\rm HS},2}&=&\displaystyle3\rho\left[
\frac{\rho(1+2\rho)(3+\rho-\rho^2)}
{\Delta_0^4}+\ln\Delta_0\right]~,\\
{f}\ex_{{\rm HS},3}&=&\displaystyle3\rho\left[
\frac{\rho(1+3\rho-2\rho^2)}
{\Delta_0^3}-\ln\Delta_0\right]~.
\eqn
Altogether we therefore have, for the perturbative expansion (\ref{evans-free}) of the
excess free energy density of the SHS mixture,
\bq \label{evans-a} 
\begin{array}{lll}
f\ex\mono&=&{f}\ex_{{\rm HS},0}+\Delta {f}\ex_0,\\
a\rho&=&{f}\ex_{{\rm HS},1}+\Delta {f}\ex_1~,\\
b\rho&=&{f}\ex_{{\rm HS},3}+\Delta {f}\ex_3~,\\
c\rho&=&{f}\ex_{{\rm HS},2}+\Delta {f}\ex_2~.
\end{array}
\eq
With these results we can now proceed to apply Evans' general results
to study cloud and shadow curves and fractionation effects in
polydisperse SHS mixtures. 

Inspection of the lengthy explicit expressions for $a$, $b$ and $c$ shows that the
dependence on the stickiness expansion coefficients $\epsilon_{1a}$,
$\epsilon_{2a}$, $\epsilon_{2b}$ is in fact rather simple. For $a$ one finds
the form 
\be
a=a_0 + \epsilon_{1a}a_1\ ,
\label{a_form}
\ee
with $a_0$ and $a_1$ functions of
$\rho$ and $\tau$ only. This is reasonable since $a$ is the coefficient of
a first order (in $\delta$) term in the excess free energy, and should
therefore only depend on the expansion of the $\epsilon_{ij}$ to the
same order. The function $b$ involves in addition terms proportional
to $\epsilon_{1a}^2$ and $\epsilon_{2b}$, while the remaining coefficient
$\epsilon_{2a}$ occurs only in the function $c$. Since $c$ does not
feature in the expressions for the phase boundary shifts or
fractionation effects to $O(s^2)$, all results we show below are
therefore independent of $\epsilon_{2a}$.

\section{Phase behavior}
\label{sec:phases}

In this section we show our results for the phase behavior of
polydisperse SHS mixtures. We will explore the various choices of
stickiness coefficients discussed in Sec.~\ref{sec:coeff}, i.e.\ 
cases I, II, IV and V as well as the AO model for small values
of the polymer-to-colloid size ratio. The first subsection has the
main results from our perturbation theory in polydispersity for the PY
closure; in Sec.~\ref{sec:comparison} we then compare these
predictions with those from other approximation schemes.

\subsection{PY closure}

\begin{figure}[ht!]
\begin{center}
\includegraphics[width=14cm]{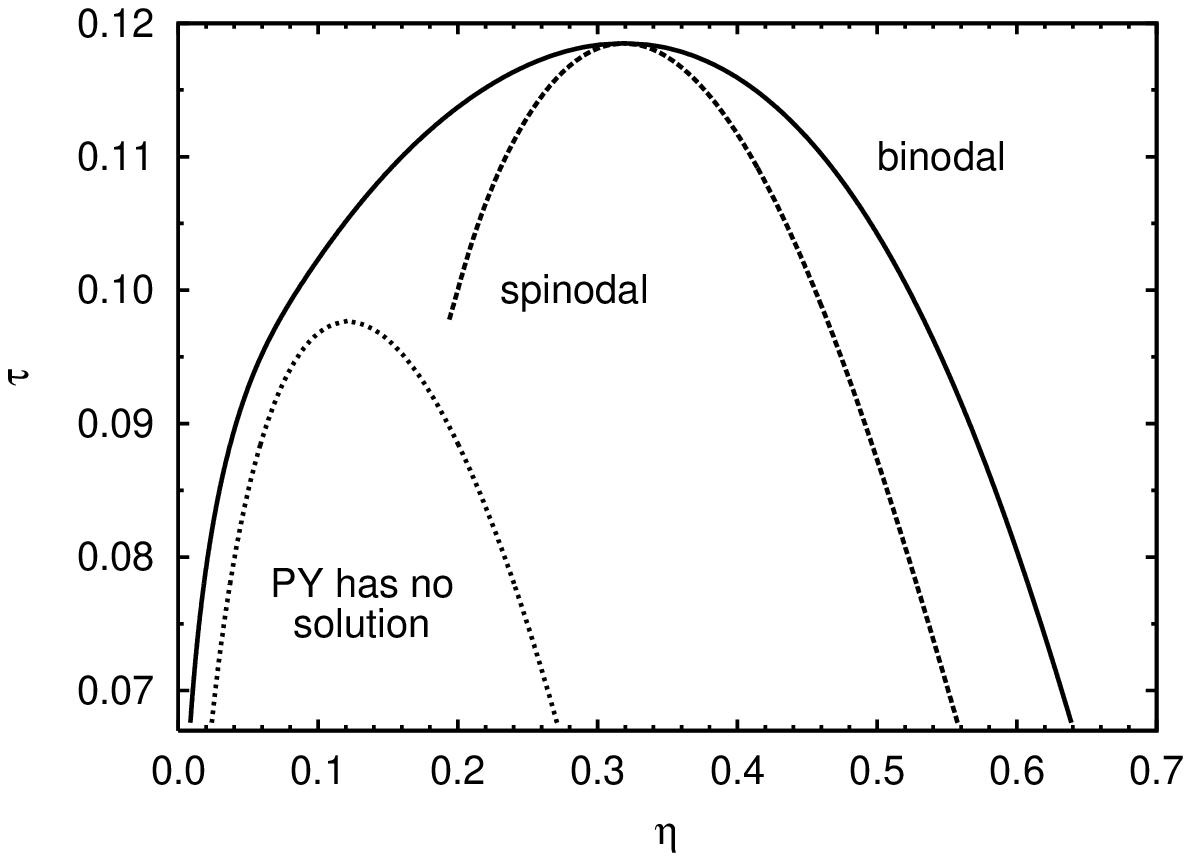}
\end{center}
\caption{}
\label{fig:pd0}
\end{figure}

We start by recalling in Fig.~\ref{fig:pd0} the phase diagram of the
monodisperse SHS fluid 
as obtained within the PY approximation and using the energy route to
thermodynamics. Along with the binodal we show the spinodal, where the
curvature of the free energy vanishes and a homogeneous phase becomes
unstable to local density fluctuations, and the
region~(\ref{existence}) where Baxter's PY equation has no physical
solution. Here and in the following we use on the $x$-axis the volume
fraction $\eta$ rather than the density $\rho$. In our units, these
two quantities are identical for monodisperse systems, but differ to
order $s^2$ in
the presence of size polydispersity. For parent phases specifically,
Eq.~(\ref{mom_expansion}) gives $\eta^\pa=\rho^\pa(1+3s^2)$ to
quadratic order. 
Cloud phases, which share the parental size distribution, have
similarly $\rho\one = \rho_0\one(1+3s^2)+\delta\rho\one$, while 
for shadow phases one finds using Eq.~(\ref{evans-frac}) that
$\rho\two = \rho_0\two[1+3(1+\Delta a)s^2] +
\delta\rho\two$~\cite{Evans01}. 

To get some initial intuition for the effects of polydispersity, it is
useful to consider first the single-phase equation of state. Fig.
\ref{fig:pr} shows plots of the dimensionless pressure against volume
fraction at several values of the polydispersity and for three choices
of the reduced temperature $\tau$. We consider here constant
stickiness coefficients (case IV) to allow a comparison with
numerical work for discrete mixtures~\cite{Robertus89}. It is
gratifying that we find qualitatively the same trend, with the
pressure decreasing with increasing 
polydispersity. Quantitatively, however, the results are not directly
comparable because in Ref.~\cite{Robertus89} the 
less accurate compressibility (rather than energy) route was used to evaluate the pressure.

To interpret physically why the pressure decreases with polydispersity
$s$ at fixed packing fraction $\eta$, we note first that such a
decrease is found also in the absence of adhesion (i.e.\ for HS). This
has been established in simulations~\cite{PhaRusZhuCha98} and is
reproduced qualitatively by the BMCSL equation of state; the intuitive
reason is that in a fluid (gas or liquid) phase a spread of
sizes allows for a more efficient packing of the particles. In such a
less ``jammed''
particle arrangement one expects to find fewer interparticle contacts
and so, in the presence of adhesion, fewer particle pairs interacting
attractively. This will {\em increase} the pressure,
counteracting the reduction, that one would expect for HS,
resulting from the more efficient 
packing. Our results are quite consistent with this: at finite $\tau$,
we find that the pressure decreases {\em less} with polydispersity
than in the HS limit $\tau\to\infty$.

The curves shown for the polydisperse cases
in Fig.~\ref{fig:pr} cannot be used to infer phase coexistence
properties directly by e.g.\ a Maxwell-construction: fractionation
means that two coexisting phases do not have properties represented by
a single relation between pressure and volume fraction. This remark
holds true quite generally for single-phase equations of state in
polydisperse systems, including e.g.\ the results obtained in Ref.~\cite{Robertus89}
within the PY compressibility route to the equation of state.
However, some more limited information on single-phase stability {\em can} be
deduced. Specifically, a single phase cannot be stable
where the pressure decreases with volume fraction. For the middle
graph of Fig.~\ref{fig:pr}, for example, where $\tau=0.1186$ is just above
the monodisperse critical point and so a monodisperse system is
still stable at all densities, the polydisperse mixtures with $s=0.2$ and
$0.3$ are already unstable in some range of densities. This means that
the region where phase separation occurs must extend to larger values
of $\tau$ for polydisperse than for monodisperse SHS, a result which
-- for case IV, as considered here -- we will find confirmed very shortly.

We next turn to explicit results for the phase behaviour, starting in
Fig.~\ref{fig:csIIIa} with cases II and IV for the
stickiness coefficients, illustrated here for parent polydispersity
$s=0.3$. The cloud curve gives the boundary of the 
region where phase coexistence occurs. The shadow curve, which records
the density of the coexisting phase at each point of incipient phase
separation, is normally distinct from this. However, for the
purely size-polydisperse cases considered here it is known on general
grounds that when represented in terms of volume fraction rather than
density the cloud and shadow curves {\em coincide} to
$O(s^2)$~\cite{WilFasSol04,Sollich06}. It is reassuring that, as
Fig.~\ref{fig:csIIIa} shows, this property is preserved by the PY
approximation.

Turning to more detailed features of Fig.~\ref{fig:csIIIa}, we
observe that in case IV the coexistence region is broadened towards
both lower and higher volume fractions. As the monodisperse critical
point is approached, the perturbation expansion breaks down as
expected and the cloud/shadow curves diverge. No quantitative
information can then be extracted in this regime, but the fact that
the divergence is {\em outwards} still tells us that the coexistence
region in the polydisperse case extends to {\em larger} values of
$\tau$ than for monodisperse SHS. This is consistent with our
inference from the single-phase equation of state above.

Comparing cases II and IV in Fig.~\ref{fig:csIIIa} one sees first that
the phase boundary shifts are rather smaller in the former than the
latter. Also the (slight) broadening of the phase separation region towards
lower $\eta$ is now restricted to $\tau$ below around 0.093, while
above the opposite trend is observed. The divergence of the curves at
the monodisperse critical point is now {\em inwards} so that phase
coexistence must terminate at a values of $\tau$ below the monodisperse $\tau_c$.

Figure \ref{fig:csIIIb} shows the cloud and shadow curves for case
V. We find that the shifts away from the monodisperse binodal are
rather larger than in the previous two cases, and therefore show
results for a smaller polydispersity $s=0.2$, rather than for $s=0.3$.
Cloud and shadow curves no longer collapse, consistent with
expectation as case V is not purely size-polydisperse. The cloud curve
shows that the coexistence region {\em narrows} in this case, except on the
high-density branch for $\tau$ below $\approx 0.085$. The inward
divergence of the cloud curve shows that the coexistence
region also shrinks towards lower $\tau$. The shadow phases are more
dense throughout than the phases on the same branch of the cloud
curve. Except for the last point, these trends agree with the
non-perturbative results
of Ref.~\cite{Fantoni05a} derived within the C0 closure.

Case I, shown in Fig.~\ref{fig:caseI}, has even stronger
polydispersity effects and we show predictions for a correspondingly
smaller polydispersity $s=0.1$. For $\tau$ not too far below the
critical point the behaviour is otherwise qualitatively similar to
case V; for lower $\tau$ the coexistence region is displaced towards
lower rather than, as in case V, higher volume fractions. The
shrinking of the coexistence region towards lower $\tau$ is again in
qualitative agreement with results from the simpler C0
closure~\cite{Fantoni05a}.


Finally we turn to the phase behaviour predicted for the AO model with
a small polymer-to-colloid size ratio $\xi/\sigma_0=0.1$ and
polydispersity $s=0.07$, as shown in Fig.~\ref{fig:csao}.  For this
choice of $\xi$ we have $\gamma\approx 3.97$ at the critical point of
the monodisperse system, and the condition $\delta_i\sim s \ll
1/\gamma$ for the validity of the expansion in particle size of the
stickiness coefficients $\epsilon_{ij}$ is reasonably well obeyed.
Here the coexistence region is broadened in all directions by the
introduction of polydispersity: towards low and high densities, and
also towards larger values of $\tau$. The shadow phases are again
more densely packed than the analogous cloud phases.

We conclude this section by considering fractionation effects. These
are illustrated in Fig.~\ref{fig:disp} for cases II and I, for a
parent distribution of Schulz form and with values of the
polydispersity $s$ as in the
corresponding Figs.~\ref{fig:csIIIa} and~\ref{fig:caseI}. When phase
separation is approached from low densities, a gas cloud phase with the
parental size distribution coexists with an infinitesimal amount of a
liquid shadow phase with a different size distribution. At the high
density boundary of the coexistence 
region, a liquid cloud phase similarly coexists with a distinct gas shadow
phase. Fig.~\ref{fig:disp} shows that for case II the liquid phase
contains more larger particles than the coexisting gas in both of
these situations (and therefore presumably throughout the whole
coexistence range of parent densities 
at the chosen $\tau$). Case I exhibits the opposite behaviour: here the
liquid phases contain more {\em smaller} particles than their
coexisting gas counterparts.

To understand this difference between cases I and II, we return to
Eq.~(\ref{evans-frac}). Consider the gas cloud point, where
$\rho\one\mono$ and $\rho\two\mono$ are the densities of coexisting
gas and liquid in the monodisperse system; $\Delta a$ then is the
difference in the values of $a$ between gas and liquid. If this is
positive, then Eq.~(\ref{evans-frac}) says that the liquid shadow has
an enhanced concentration of larger particles. By reversing the role
of the two densities one then sees easily that also at the liquid
cloud point the liquid phase will contain more of the larger particles
than the gas 
(shadow) phase. In summary, the liquid contains predominantly the larger particles
if $\Delta a>0$, and the smaller particles if $\Delta a<0$. But from
Eq.~(\ref{a_form}), $\Delta a = \Delta a_0 + \epsilon_{1a}\Delta a_1$
so that different choices of stickiness coefficients affect the
direction and strength of fractionation only via $\epsilon_{1a}$. The
functions $\Delta a_0$ and $\Delta a_1$ are shown in
Fig.~\ref{fig:delta_a} and are both positive; as a result, $\Delta a$
is positive when $\epsilon_{1a}>-\Delta a_0/\Delta a_1$ and negative
otherwise. The ratio occurring on the r.h.s.\ is almost constant and remains
close to 1/3 over a large range of $\tau$, as the inset of
Fig.~\ref{fig:delta_a} demonstrates. We can now rationalize the
difference between cases I and II observed above: for case I,
$\epsilon_{1a}=-1<-1/3$, hence $\Delta a<0$ and fractionation will
enrich the liquid in small particles; for case II,
$\epsilon_{1a}=0>-1/3$ and one has the opposite situation. Referring
to Table~\ref{tab:peps} we also conclude that case IV will have
the same fractionation behaviour as case II, while case V will produce
the same ``direction'' of fractionation (smaller particles in the
liquid) as case I but with quantitatively weaker effects. In the AO case
$\epsilon_{1a}$ depends on $\tau$ as discussed in Sec.~\ref{sec:AO} but
this effect turns out
to be weak quantitatively, with (for $\xi/\sigma_0=0.1$) $\epsilon_{1a}$ ranging from
$\approx 0.95$ at the critical point to $\approx 1.24$ at
$\tau=0.065$. Taking for simplicity $\epsilon_{1a}\approx 1$ one
infers that fractionation effects will be qualitatively similar to
cases II and IV, but quantitatively $\Delta a$ will be larger by a
factor of around $4$. All of these
conclusions can be confirmed by detailed examination of the explicit results
for the various cases. 

\subsection{Other approximation schemes}
\label{sec:comparison}

Once one accepts the PY closure, the results shown above are exact in
their treatment of polydispersity, certainly within the perturbative
setting of weakly polydisperse mixtures. However, the PY closure
itself -- while more accurate than its competitors -- does
remains an approximation. It is therefore useful to compare with the
predictions of other approximation schemes to assess the robustness of
our predictions. We do this first for case II, where an approximate
free energy of BMCSL type can be constructed, and then for the AO model,
which can be analysed using the free volume theory of
Refs.~\cite{LekPooPusStrWar92,FasSol05}.

To construct the alternative approximation for case II one starts
from a virial expansion of the excess free energy density up to the
third virial coefficient. This is easily found as
\bq \nonumber
f\ex&=&\rho\mom_3+(3-12t)\mom_1\mom_2+\\ \label{f2ex3}
&&\frac{1}{2}\left[\rho\mom_3^2+3(1-12t+48t^2-32t^3)\mom_2^3+
6(1-4t)\mom_1\mom_2\mom_3\right]~,
\eq
where $t=1/(12\tau)$; the terms of second order in density agree
with the energy route of the C0 approximation~\cite{Fantoni05a}. The
interesting feature of this result is that the fourth order moment
$\mom_4$ does not appear, in contrast to the analogous expansions for
the other cases I, IV and V that we have considered. Furthermore, the
only modification compared to the pure HS 
case is in the $t$-dependence of the coefficients. These observations
suggest that it should be possible to construct a
modified free energy expression of BMCSL-type which matches the above virial expansion 
to third order in density. Remarkably, if the desired modified
BMCSL form is parametrized in a fairly general manner as
\bq \label{f2ex3bmcsl}
f\ex &=&\left(A_1\frac{\mom_2^3}{\mom_3^2}-A_2\rho\right)
\left[\ln(1-D\mom_3)+E\right]+\frac{3B\mom_1\mom_2}{1-D\mom_3}+
\frac{C\mom_2^3}{\mom_3(1-D\mom_3)^2}~,
\eq 
then by expanding to third order in density and matching to
the expansion (\ref{f2ex3}) one finds a {\em unique} solution for the
coefficients:
\bqn
E=0~,~~~D=A_2=1~,~~~B=1-4t~,~~~C=A_1=B^3+32t^3~.
\eqn
The presence of polydispersity is crucial here: for a monodisperse
system, the matching conditions to third order in density would not
constrain the coefficients sufficiently. 

One can now apply the perturbative scheme used throughout this paper
to obtain from the excess free energy of Eq.~(\ref{f2ex3bmcsl}) the
functions $a$ and $b$, and hence the cloud and shadow curves. (Note
that the perturbative approach is used here mainly for ease of
comparison with our other results; since the free
energy~(\ref{f2ex3bmcsl}) is truncatable, a full solution of the phase
equilibrium conditions would be fairly straightforward.) The results
are shown in Fig.~\ref{fig:csb3}; note that not just the polydisperse
cloud/shadow curves but also the monodispere binodal are different
from the ones obtained from the PY approximation. Looking at the
polydispersity-induced shifts, one sees that on the high-density
branch of the cloud/shadow curve these are quite comparable to those
from the PY approximation (Fig.~\ref{fig:csIIIa}), even
semi-quantitatively. Polydispersity effects on the low-density branch
are rather smaller, again as found within the PY closure. Near the
critical point, however, the trends are reversed: the BMCSL-type
approximation predicts an extension of the coexistence region towards
larger $\tau$ and smaller $\eta$, whereas the PY approximation leads
to the opposite result.

The second case where we have an alternative approximation scheme
available for comparison is the AO model. The free volume theory of
Ref.~\cite{LekPooPusStrWar92} effectively linearizes the excess free
energy in the polymer (reservoir) potential $\rho_p$, and the same is
true for its generalization to polydisperse
colloids~\cite{FasSol05}. It is therefore most accurate when the
depletion interaction between the colloids, which is proportional to
$\rho_p$, is small (in units of $k_{\rm B}T$). In order to still get
gas-liquid phase separation, the polymer size $\xi$ must then not be
too small. This is the opposite limit as for our SHS mapping, which
will work best when $\xi\ll\sigma_0$ and the depletion attraction is
large at contact. If anything one therefore expects the best agreement
between the two approximations for intermediate values of $\xi$; a
suitable choice is $\xi/\sigma_0=0.1$ as investigated
above. Fig.~\ref{fig:free_vol} compares the two sets of cloud and
shadow curves predicted. On the vertical axis we plot the polymer
(reservoir) volume fraction $\eta_p$. This equals $\rho_p\xi^3$ in our
dimensionless units and is the conventional variable used in phase
diagrams of colloid-polymer
mixtures~\cite{LekPooPusStrWar92}.
Comparison of the two panels of
Fig.~\ref{fig:free_vol} reveals that the qualitative agreement between
the two theories is surprisingly good. In particular, the qualitative
changes caused by the presence of polydispersity (broadening of
coexistence region to lower and higher colloid volume fraction and
lower polymer volume fraction) are in full agreement. For the relevant
range of polymer volume fractions there is even quite good
quantitative agreement (though note the slightly different axis ranges
on left and right), and the shifts of cloud and shadow curves away
from the monodisperse binodal are also quite comparable. Even the
predicted fractionation effects agree well: as the inset on
the right of Fig.~\ref{fig:free_vol} demonstrates the calculated
values of $\Delta a$ are, apart from the slight shift in the critical
point values of the polymer volume fraction, quite consistent with
each other.

We note briefly that in order to calculate the free volume theory
data shown in Fig.~\ref{fig:free_vol} we took the excess free energy
for fully polydisperse colloids (at fixed polymer chemical potential)
derived in Ref.~\cite{FasSol05} and then found the functions $a$ and $b$ by
expanding explicitly as in Eq.~(\ref{evans-free}). This gives for
$a$ the same result as obtained by Evans~\cite{Evans01}, while $b$
differs from his expression in terms of approximate correlation
functions~\cite{Evans01}. One might expect that our approach of deriving $a$ and $b$
from one unified polydisperse excess free energy would be somewhat
more accurate than Evans' procedure of finding $a$ and $b$ by quite
different routes. We have checked that for larger polymer sizes
$\xi/\sigma_0=0.4$ our method predicts similar trends to those reported
in~\cite{Evans01}, but quantitatively the effects of polydispersity
are less pronounced.

\section{Conclusions}
\label{sec:conclusions}

We have presented a perturbative approach to the determination of the
gas-liquid phase behaviour of
polydisperse Sticky Hard Spheres (SHS), studied within the Percus
Yevick (PY) integral equation theory. For arbitrary size
polydispersity, the calculation of phase diagrams analogous to those
reported here would normally require the solution of a large (or infinite)
system of quadratic coupled equations, a task which in practice can be
accomplished neither analytically nor numerically. To get around this
bottleneck of the PY closure we focussed on weakly
polydisperse mixtures, where the overall size distribution is narrow in the
sense that its normalized (by the mean) standard deviation $s$ is
small compared to unity. This allowed us to calculate in closed form
the leading order ($O(s^2)$) shifts of cloud and shadow curves away
from the monodisperse binodal, and the corresponding fractionation
effects. The thermodynamics was derived from the PY solution via the
energy route because in the monodisperse case this method gives the best
match to Monte Carlo simulation results, even for low reduced
temperatures $\tau$ around and below the critical point.

In order to specify the properties of a SHS mixture one needs to know
how the stickiness coefficients $\epsilon_{ij}$ depend on the sizes of
the two interacting particles. We discussed a number of plausible
constraints on this size dependence. In obtaining explicit results we
considered specifically the cases I-V (excluding III, which with
our now more appropriate choice of reference length becomes identical
to I) previously suggested within exact solutions of simpler closures
like C0 and C1. Of these, cases II and IV are special since they can be
seen as the sticky limit of purely size-polydisperse interactions, in
which scaling of both particle sizes by a common factor only changes
the range but not the 
strength of the interaction. We have also considered the AO model of a
mixture of polydisperse colloids and polymers, which for small polymer
size can be mapped to a good approximation onto an SHS model. The
stickiness coefficients can be derived in this case rather than postulated; in
contrast to the simpler ad hoc prescriptions of cases I-V, they are
functions of $\tau$.


In the simplest case IV of constant stickiness coefficients we first
investigated the single-phase equation of state, finding qualitative
agreement with a numerical solution of the compressibility equation of
state for a small number of components by Robertus et
al.~\cite{Robertus89}. Moving on to phase coexistence proper, 
we found for cases II and IV that cloud and shadow curves coincide in
in the volume fraction representation and to $O(s^2)$, as expected on
general grounds; less obviously our results also show that in these
two cases the deviations of the polydisperse cloud/shadow curves away from
the monodisperse binodal are quantitatively small. In all the other
cases considered the shadow curves are located at higher
volume fractions than the cloud curves, a trend observed in many other
polydisperse systems~\cite{poly_review,Evans01}.

Summarizing our findings regarding the effect of polydispersity on the
extent of the coexistence region as delimited by the cloud curve, it
is simplest initially to group the different scenarios according to
their behaviour near the critical point. For case IV and the AO model
(with a polymer-to-colloid size ratio of 0.1) the coexistence region
is shifted to higher reduced temperatures $\tau$; conversely, at fixed
$\tau$ it covers a wider range of parent volume fractions
$\eta$. Cases I, II and V, on the other hand, show the opposite
behaviour, with the coexistence region shrinking towards lower $\tau$.

The trends in cases IV and AO remain unchanged as one moves to lower
values of $\tau$, with the coexistence region continuing to broaden
towards lower and higher values of $\eta$ at the two ends (gas and
liquid). In the other cases the shrinking trend near the critical
point can be reversed at lower $\tau$. E.g.\ for case II one also
eventually sees a broadening to lower (gas branch) and higher (liquid
branch) $\eta$. For case V the coexistence region is shifted to {\em
higher} $\eta$ at both ends (gas and liquid) at low $\tau$; case I
shows the opposite behaviour.

We have analyzed also the fractionation effects that accompany
polydisperse phase separation, where coexisting phases have different
particle size distributions. Depending on the stickiness
coefficients considered, the liquid phase contains predominantly
the larger (as in cases II, IV and AO) or the smaller particles (as in
cases I and V). We rationalized this result by showing that the fractionation
effects depend on the stickiness coefficients only via the expansion coefficient
$\epsilon_{1a}$; where this is above $\approx -1/3$, the larger
particles accumulate in the liquid phase, otherwise in the gas.

Finally we have compared our results with the predictions from other
available approximation schemes, to check their robustness. Case II is
important here because a variety of simple but realistic interactions
potentials, used in the 
literature to model short ranged attractions in real solutions of
colloids, reverse micelles or globular proteins, can be mapped onto
this model~\cite{Gazzillo06}. We constructed an approximate excess
free energy by allowing various coefficients within the BMCSL free
energy for hard spheres to be come $\tau$-dependent and matching to
the (for case II, particularly simple) third order virial expansion.
The resulting binodal in the monodisperse limit is rather different
from the one obtained from the PY closure with the energy route. The
polydispersity-induced shifts of the (coincident) cloud/shadow curves
are nevertheless comparable to those predicted by our PY analysis, but
only sufficiently far below the critical point. Near the critical
point the BMCSL-like excess free energy predicts an enlargement of the
coexistence region towards higher $\tau$, while the PY closure gives
the opposite result. Given that in the monodisperse case the PY
binodal is rather closer to simulation results than the BMCSL-like
one, we would expect that also for the polydispersity effects the PY
predictions are more accurate.

The second model for which we considered an alternative approximation
scheme was the AO model. Here a direct comparison with free volume
theory is straightforward since for the latter a generalization to
polydisperse colloids has recently been derived~\cite{FasSol05}. Even
though one expects the two 
approaches to be valid in complementary regions (small polymer size
$\xi$ for the SHS mapping, larger $\xi$ for free volume theory) we
found very good qualitative and even semi-quantitative agreement of
the predictions from the two routes for an intermediate value (0.1) of
the polymer-to-colloid size ratio.

In future work, direct simulations of polydisperse SHS mixtures would
obviously be of interest to test our predictions and resolve any
differences with other approximation schemes, e.g.\ in case
II. Simulations would be ideal here since in contrast to experiment
they would allow one to probe directly different choices for the
stickiness coefficients. Because of the presence of polydispersity, a
grand canonical Monte Carlo
approach~\cite{WilFasSol04,WilSol02,WilSol04,WilSolFas05} may be the
simulation method of choice, possibly supplemented by specific cluster
algorithms tailored to sticky
interactions~\cite{Miller03,Miller04a,Miller04b}. For the physically
more realistic AO model, our
predictions should be more accurate than those of free volume theory
for small polymer-to-colloid size ratios. Detailed experimental or
simulation tests in this regime would be welcome. In simulations one
could work directly with the AO-depletion potential for the colloids,
without ever representing the polymers explicitly.  For comparison
with experiment one would need to work out the actual volume fraction
of polymer in the system rather than in a reservoir; this should in
principle be a straightforward exercise once our excess free energy
has been rewritten as a function of polymer chemical potential. On the
experimental side one would require that the colloids are sufficiently
polydisperse (beyond a terminal polydispersity around $s=0.07$; see
the discussion and bibliography in Ref.~\cite{FasSol04}) to suppress
kinetically any solid phases, thus allowing 
stable observation of the gas-liquid phase splits we have
calculated.

\appendix
\section{Perturbative expansion of $L_{ij}$}
\label{app:a}

For the perturbative expansion of Eq.~(\ref{pss}) one needs the
expansions of $\alpha_{ij}$, $\beta_{ij}$ and $\phi_{ij}$. These
involve the trivial expansions
\bq \label{sigi1}
\sigma_i&=&1+\delta_i~,\\
\sigma_{ij}&=&1+\frac{1}{2}(\delta_i+\delta_j)~,\\
\label{sigi3}
\sigma_i\sigma_j&=&1+(\delta_i+\delta_j)+\delta_i\delta_j~.
\eq
One also needs the expansions to quadratic order of the moments
\be
\mom_m=\rho\langle(1+\delta)^m\rangle =
\rho(1+m\avdel+\frac{1}{2} m(m-1)\avdels+\ldots)~,
\label{mom_expansion}
\ee
giving in particular $\mom_2
= \rho(1+2\avdel+\avdels)$ and
$\Delta=1-\eta=1-\rho_3=\Delta_0-3\rho\avdel-3\rho\avdels$ with
$\Delta_0=1-\rho$ as defined in the main text. The final ingredient is
the expansion (\ref{epsij_expansion}) for the $\epsilon_{ij}$, which
is left in general form to allow different possible choices of the
stickiness coefficients to be considered together. Altogether one gets
the following expansion coefficients for the $\alpha_{ij}$:
\bqn
\begin{array}{lll}
\alpha_0\tau&=&\frac{1}{\Delta_0}
+\frac{3}{2}\frac{\rho}{\Delta_0^2}~,\\
\alpha_{1a}\tau&=&(1+\epsilon_{1a})\frac{1}{\Delta_0}
+\left(\frac{9}{4}+\frac{3}{2}\epsilon_{1a}\right)
\frac{\rho}{\Delta_0^2}~,\\
\alpha_{1b}\tau&=&6\frac{\rho}{\Delta_0^2}
+9
\frac{\rho^2}{\Delta_0^3}~,\\
\alpha_{2a}\tau&=&\left(\frac{1}{2}+2\epsilon_{1a}+
\epsilon_{2a}\right)\frac{1}{\Delta_0}
+\left(3+\frac{9}{2}\epsilon_{1a}+\frac{3}{2}\epsilon_{2a}\right)
\frac{\rho}{\Delta_0^2}~,\\
\alpha_{2b}\tau&=&\left(\frac{1}{4}+\epsilon_{1a}+
\epsilon_{2b}\right)\frac{1}{\Delta_0}
+\left(\frac{3}{4}+\frac{9}{4}\epsilon_{1a}+\frac{3}{2}
\epsilon_{2b}\right)\frac{\rho}{\Delta_0^2}~,\\
\alpha_{2c}\tau&=&
\left(\frac{15}{2}+6\epsilon_{1a}\right)
\frac{\rho}{\Delta_0^2}
+\left(\frac{27}{2}+9\epsilon_{1a}\right)
\frac{\rho^2}{\Delta_0^3}~,\\ 
\alpha_{2d}\tau&=&27\frac{\rho^2}{\Delta_0^3}
+\frac{81}{2}\frac{\rho^3}{\Delta_0^4}~,\\
\alpha_{2e}\tau&=&
\frac{9}{2}\frac{\rho}{\Delta_0^2}
+9\frac{\rho^2}{\Delta_0^3}~.
\end{array}
\eqn
Similarly one has for the $\beta_{ij}$
\bqn
\begin{array}{l}
\beta_0\tau=\rho~,\\
\beta_{1a}\tau=\left(\frac{1}{2}+\epsilon_{1a}\right)
\rho~,\\
\beta_{1b}\tau=0~,\\
\beta_{2a}\tau=\left(\epsilon_{1a}+\epsilon_{2a}\right)
\rho~,\\
\beta_{2b}\tau=\left(\frac{1}{2}\epsilon_{1a}+\epsilon_{2b}\right)
\rho~,\\
\beta_{2c}\tau=0~,\\
\beta_{2d}\tau=0~,\\
\beta_{2e}\tau=0~,
\end{array}
\eqn
and for the $\phi_{ij}$
\bqn
\begin{array}{l}
\phi_0=\frac{1}{\Delta_0}~,\\
\phi_{1a}=\frac{1}{\Delta_0}~,\\
\phi_{1b}=\frac{3\rho}{\Delta_0^2}~,\\
\phi_{2a}=\frac{1}{\Delta_0}~,\\
\phi_{2b}=0~,\\
\phi_{2c}=\frac{3\rho}{\Delta_0^2}~,\\
\phi_{2d}=\frac{9\rho^2}{\Delta_0^3}~,\\
\phi_{2e}=\frac{3\rho}{\Delta_0^2}~.
\end{array}
\eqn

One now inserts these expansions into Eq.~(\ref{pss}) and proceeds as
explained in the main text to obtain
the desired conditions on the expansions coefficients
$L_0,\ldots,L_{2e}$ of the $L_{ij}$.
To state these, it is helpful to define the quantities
\bqn
M_{\alpha,\beta}\equiv\frac{1}{12}L_\alpha L_\beta-\frac{1}{2}
(L_\alpha\phi_\beta+L_\beta\phi_\alpha)~,
\eqn
where Greek indices stand for the labels 0, 1a, 1b, 2a, 2c, 2d, 2e,
of the coefficients of the perturbative expansions. The desired
conditions are then
\bq \label{o0}
L_0&=&\alpha_0+\beta_0M_{0,0}~,
\\
\label{o1a}
L_{1a}&=&\alpha_{1a}+\beta_{1a}M_{0,0}+\beta_0M_{0,1a}~,
\\
\label{o1b}
L_{1b}&=&\alpha_{1b}+\beta_{1b}M_{0,0}+2\beta_0(M_{0,1a}+M_{0,1b})~,
\\
\label{o2a}
L_{2a}&=&\alpha_{2a}+\beta_{2a}M_{0,0}+2\beta_{1a}M_{0,1a}+
\beta_0M_{1a,1a}~,
\\
\label{o2b}
L_{2b}&=&\alpha_{2b}+\beta_{2b}M_{0,0}+\beta_{1a}M_{0,1a}
+\beta_0M_{0,2b}~,
\\
\nonumber
L_{2c}&=&\alpha_{2c}+\beta_{2c}M_{0,0}+
2\beta_{1a}(M_{0,1a}+M_{0,1b})+
\beta_{1b}M_{0,1a}\\ \label{o2c}
&&{}+{}\beta_{0}(M_{1a,1a}+M_{1a,1b}+M_{0,2a}+M_{0,2c})~,
\\
\nonumber
L_{2d}&=&\alpha_{2d}+\beta_{2d}M_{0,0}
+2\beta_{1b}(M_{0,1a}+M_{0,1b})
\\
\label{o2d}
&&{}+{}\beta_{0}(2M_{0,2c}+2M_{0,2d}+2M_{1a,1b}+M_{1b,1b})~,
\\
\label{o2e}
L_{2e}&=&\alpha_{2e}+\beta_{2e}M_{0,0}
+\beta_{0}(M_{1a,1a}+2M_{0,2b}+2M_{0,2e})~,
\eq
The first of these determines $L_0$ and leads back to Baxter's
solution (\ref{L0}) for the monodisperse case. All other equations
involve the desired coefficient on the left at most linearly on the
right hand side and so are trivial to solve; e.g.~Eq.~(\ref{o1a}) has
$L_{1,a}$ on the left and implicitly via $M_{0,1a}$ on the right.
Running through the equations in order, all expansion coefficients can
then be found.

\bibliographystyle{apsrev}
\bibliography{shspypp11}

\begin{thebibliography}{48}
\bibitem[{not()}]{note2}
\bibinfo{note}{Note that the energy route to the equation of state of HS fluids
  is ill defined since the internal energy is just that of the ideal gas. So in
  the PY approximation one could use \cite{Salacuse82} either the
  compressibility pressure $P^{\rm PY(c)}$ or the virial pressure $P^{\rm
  PY(v)}$. For a mixture of a discrete number of species it is known that the
  BMCSL pressure $P^{\rm BMCSL}=(2P^{\rm PY(c)}+P^{\rm PY(v)})/3$ is more
  accurate than either of these two limits. Recently, the equivalence between
  the energy and virial routes to the equation of state for the one-component
  HS fluid has been shown in \cite{Santos05}. We use the BMCSL expression
  throughout, except in figure 1 where the PY compressibility pressure is used
  for ease of comparison with earlier work~\cite{Fantoni05a}}.
\end{thebibliography}
\newpage
\centerline{\bf LIST OF FIGURES}

\begin{itemize}
\item[Fig. 1] Equation of state, from the energy route, for a one-component
fluid of SHS. From left to right and top to bottom the four panels
refer respectively to a reduced temperature of $\tau=1.00,0.50,0.20,$ and
$0.15$. The continuous line corresponds to the MSA approximation, the
dotted line to the mMSA approximation, the short dashed line to the C1
approximation, the long dashed line to the PY approximation, the
dot-dashed line to the WCA first order perturbation theory, squares to
the WCA second order perturbation theory (with error bars indicating
the range where the true value should lie with probability 99.7\%),
and triangles to the MC simulations of Miller and Frenkel
\cite{Miller04a}. In all cases the HS component of the pressure was
chosen to be the one obtained from the compressibility route of the PY
approximation \cite{note2}.

\item[Fig. 2] The overlap volume ${\cal V}_{\rm ov}(r)$ of the 
two exclusion zones around colloid particles of diameter $\sigma_i$ and 
$\sigma_j$ which cannot be accessed by polymers of diameter $\xi$.

\item[Fig. 3] Phase diagram of the monodisperse SHS fluid obtained with
the PY closure and the energy route to thermodynamics. Shown are the 
binodal and spinodal curves and the region where the PY equation has
no solution (see Eq.~(\ref{existence})).

\item[Fig. 4] Pressure from the energy route of the PY approximation for
a single (parent) phase with case IV stickiness
coefficients, plotted against volume fraction. Results are shown for
several small values of the  
polydispersity $s$ (see legend) and well above, just above, and below
(from left to right) the 
critical point of the monodisperse system. The pressure was determined
using Eq.~(9) of Ref.~\cite{Evans01}.

\item[Fig. 5] Cloud and shadow curves for 
SHS mixtures with polydispersity $s=0.3$, as obtained
within the PY
approximation and the energy route to thermodynamics, for
coefficients $\epsilon_{ij}$ chosen according to cases II and IV from
Eq.~(\ref{Cases}). The shifts from 
the binodal of the monodisperse system (labeled ``mono'') were
calculated using Eq.~(\ref{evans-shift})
and give the leading ($O(s^2)$) corrections in a perturbative
treatment of polydispersity. Note the collapse of the
cloud and shadow curve, as expected to this order of the perturbation
theory for purely size-polydisperse models~\cite{WilFasSol04,Sollich06},
and the divergence of the perturbation theory at the monodisperse
critical point.

\item[Fig. 6] Cloud and shadow curves for the SHS model with polydispersity
$s=0.2$ and case V stickiness coefficients. The binodal of the monodisperse 
system is shown for comparison.

\item[Fig. 7] Cloud and shadow curves for the SHS model with polydispersity
$s=0.1$ and case I stickiness coefficients. The binodal of the monodisperse 
system is shown for comparison.

\item[Fig. 8] Cloud and shadow curves for the AO model with
polymer-to-colloid size ratio $\xi/\sigma_0=0.1$ and (colloid)
polydispersity $s=0.07$. 
The binodal of the monodisperse system is shown for comparison.

\item[Fig. 9] Fractionation in SHS mixtures with stickiness coefficients
chosen according to cases II and I, at $\tau=0.11$ and for polydispersities 
$s$ as in the corresponding Figs.~\ref{fig:csIIIa}
and~\ref{fig:caseI}. Shown are the cloud (parent) size distribution
$p(\sigma)$, taken to be of Schulz form, and the size distributions
in the liquid shadow and gas shadow phases that form when coexistence 
is approached from low densities (gas cloud phase)
and high densites (liquid cloud phase), respectively. For case II
(main graph) the larger particles tend to accumulate in the liquid phase, while for
case I (inset) the opposite is true.

\item[Fig. 10] Decomposition $\Delta a = \Delta a_0+\epsilon_{1a}\Delta a_1$
of the difference in $a$ between gas and liquid phases. The two
contributions $\Delta a_0$ and $\Delta a_1$ are plotted separately
against $\tau$; the latter quantity is graphed on the vertical
rather than the horizontal axis for ease of comparison with
Figs.~\ref{fig:csIIIa} to~\ref{fig:csao}. Inset: Ratio $\Delta
a_0/\Delta a_1$.

\item[Fig. 11] Cloud and shadow curves for case II stickiness coefficients
and with polydispersity $s=0.3$, calculated using the BCMSL-type free
energy, Eq.~(\ref{f2ex3bmcsl}), rather than the PY approximation as in
Fig.~\ref{fig:csIIIa}. The binodal of the monodisperse system, which
differs from the PY result, is shown
for comparison. Main graph: region
around the critical point. Inset: global view of the results on the
same scale as in Fig.~\ref{fig:csIIIa}.

\item[Fig. 12] Comparison of predictions for AO model with
polymer-to-colloid size ratio $\xi/\sigma_0=0.1$. Left: Results of SHS
mapping analysed within PY approximation; as in Fig.~\ref{fig:csao}
cloud and shadow curves are shown for colloid polydispersity
$s=0.07$, along with the monodisperse binodal for comparison. The
vertical axis now shows the polymer volume fraction rather than the
reduced temperature $\tau$. Right: Analogous results obtained from
free volume theory. Inset right: Fractionation coefficient $\Delta a$ for
the two approximation schemes.

\end{itemize} 
\newpage

\centerline{\bf LIST OF TABLES}
\begin{itemize}

\item[Tab. 1] Coefficients of the perturbative expansion
(\ref{epsij_expansion}) of the adhesion parameters $\epsilon_{ij}$ for
the four cases listed in Eq.~(\ref{Cases}). 

\end{itemize} 
\newpage

\begin{table}[h!]
\begin{center}
\begin{tabular}{lccccccccccc}
\hline
&$~~~~~~~$& Case I &$~~~~~~~$& Case II &$~~~~~~~$& Case IV
&$~~~~~~~$& Case V  \\
\hline
$\epsilon_0   $   && 1   && 1    && 1 && 1     \\
$\epsilon_{1a}$   && -1  && 0    && 0 && -1/2  \\
$\epsilon_{2a}$   && 3/2 && 1/2  && 0 && 1/2   \\
$\epsilon_{2b}$   && 3/4 && -1/4 && 0 && 1/4   \\
\hline
\end{tabular}
\caption[]{}
\label{tab:peps}
\end{center}
\end{table}

\newpage
%
%
%
%
%
%
%
%
\begin{figure}[ht!]
\begin{center}
\includegraphics[width=14cm]{pressure.eps}
\end{center}
\caption{} 
\label{fig:pr}
\end{figure}
%
%
\begin{figure}[ht!]
\begin{center}
\includegraphics[width=14cm]{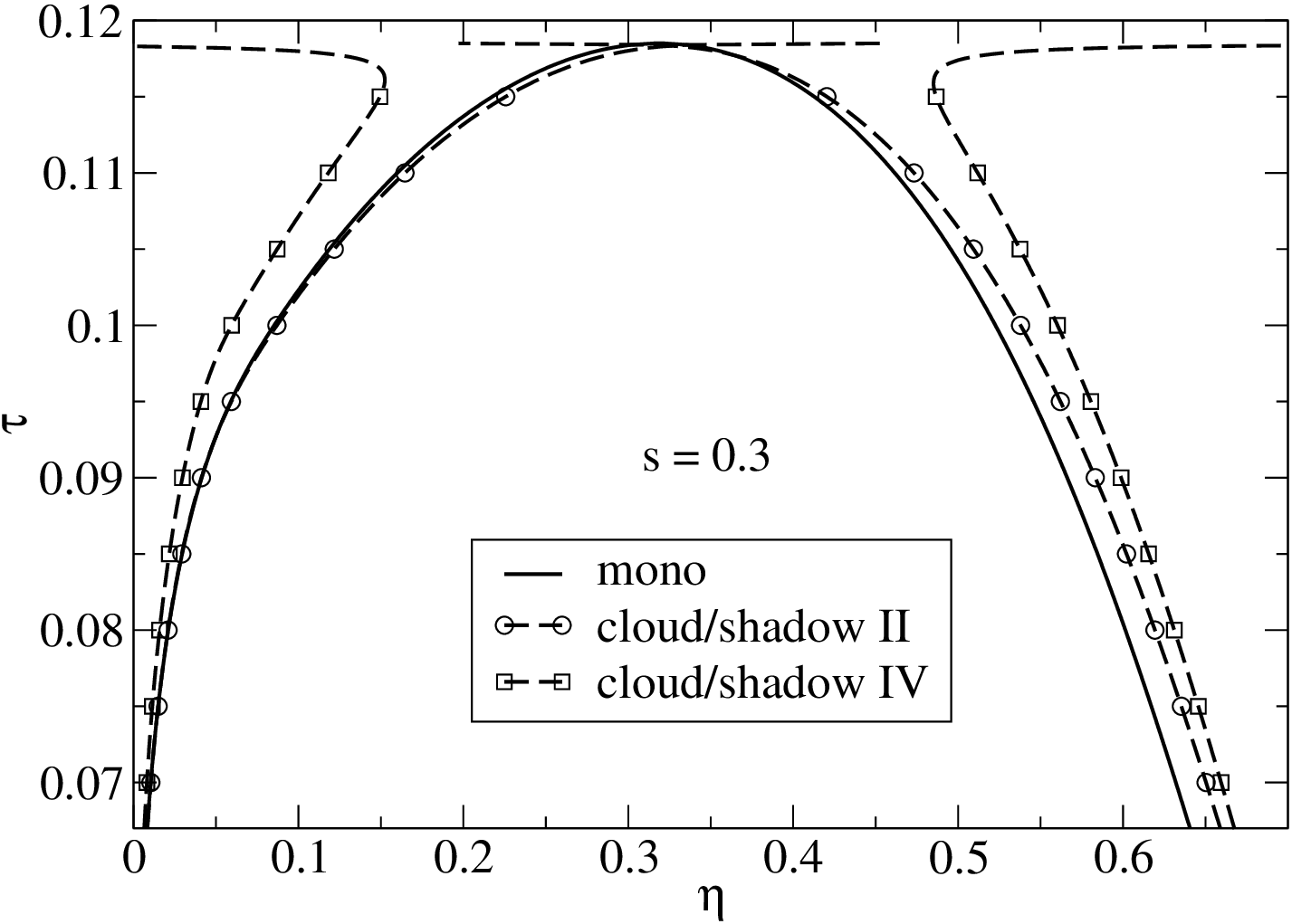}
\end{center}
\caption{}
\label{fig:csIIIa}
\end{figure}
%
%
\begin{figure}[ht!]
\begin{center}
\includegraphics[width=14cm]{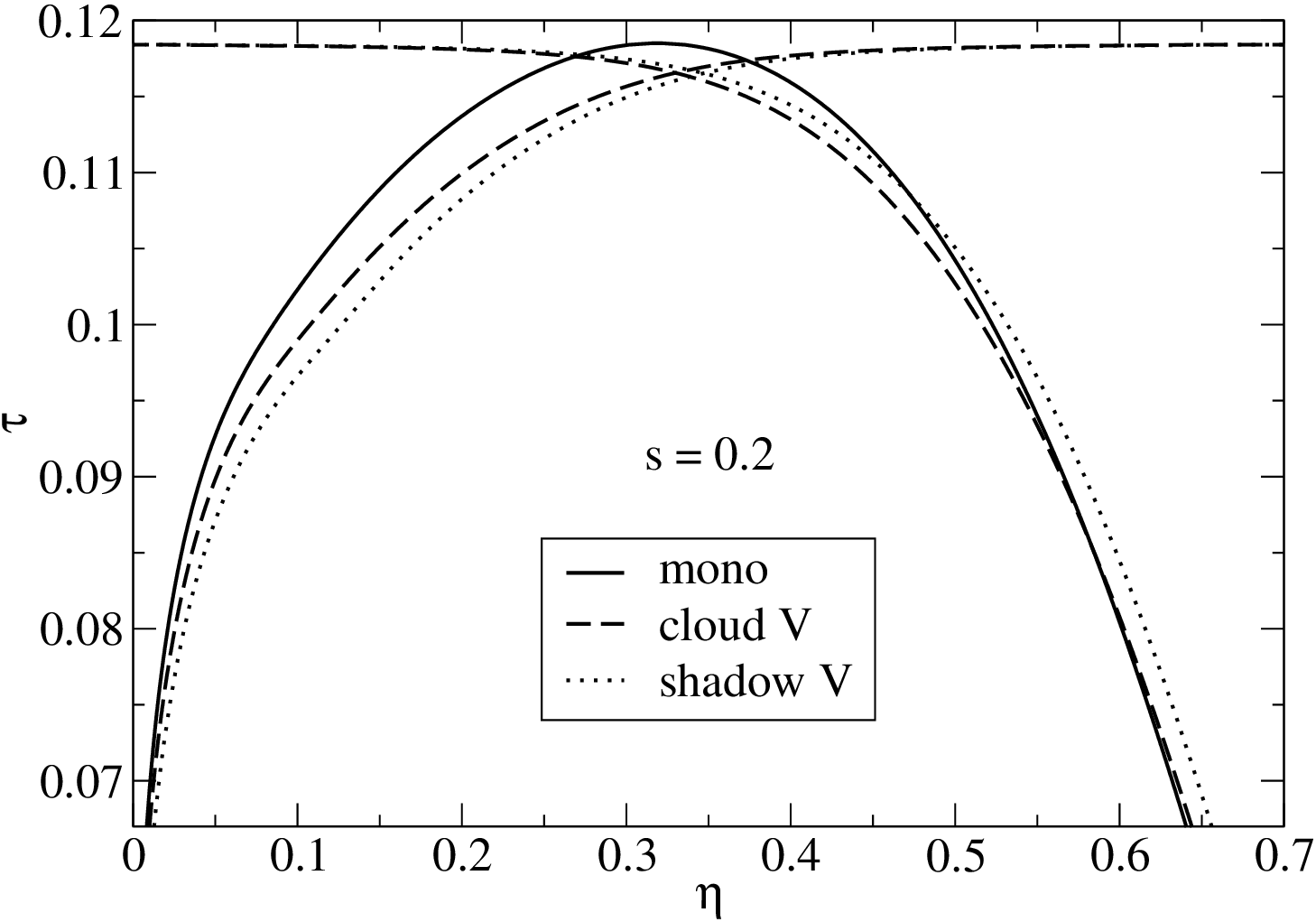}
\end{center}
\caption{} 
\label{fig:csIIIb}
\end{figure}

%
\begin{figure}[ht!]
\begin{center}
\includegraphics[width=14cm]{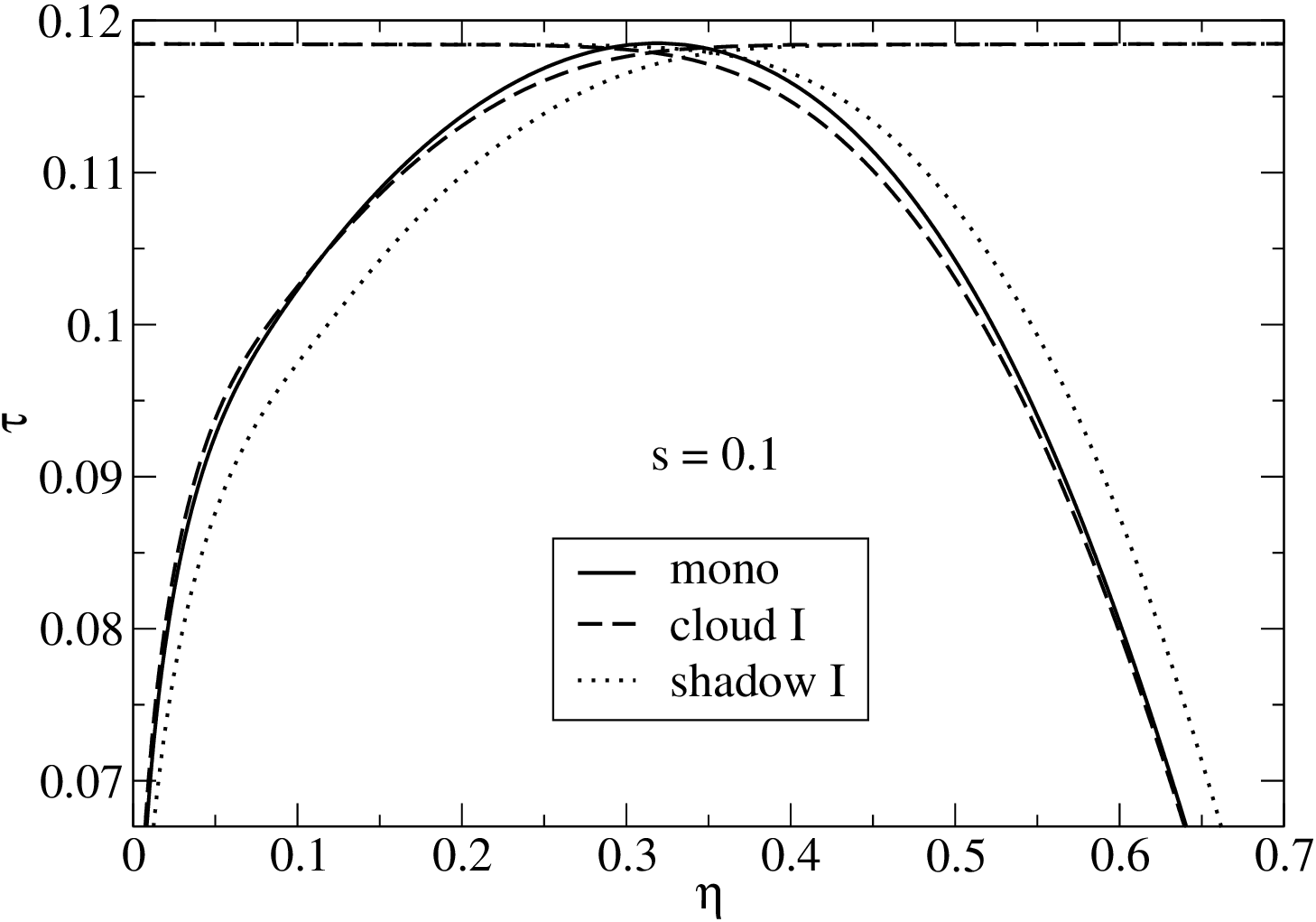}
\end{center}
\caption{}
\label{fig:caseI}
\end{figure}
%
%
\begin{figure}[ht!]
\begin{center}
\includegraphics[width=14cm]{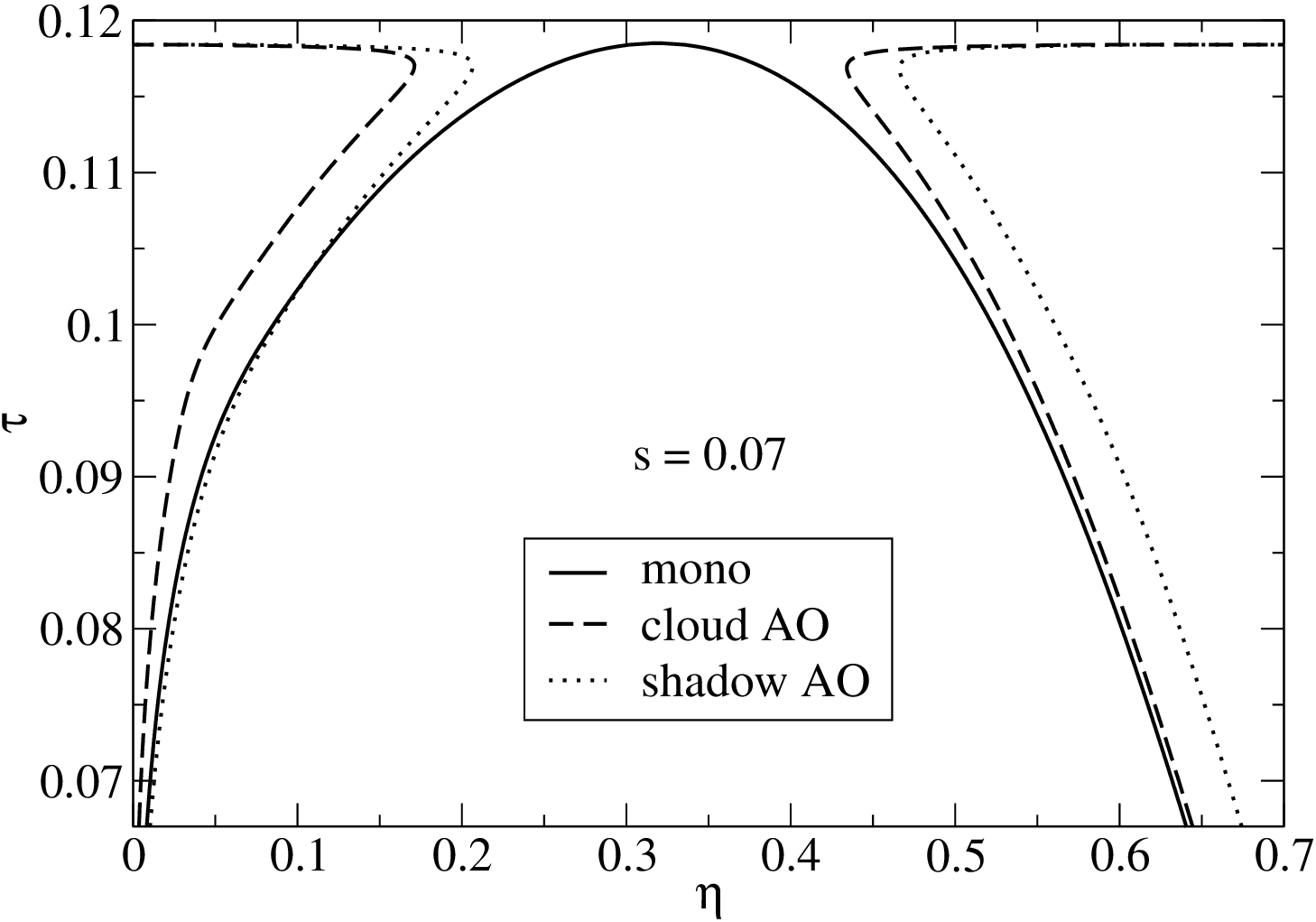}
\end{center}
\caption{} 
\label{fig:csao}
\end{figure}
%
%
\begin{figure}[ht!]
\begin{center}
\includegraphics[width=14cm]{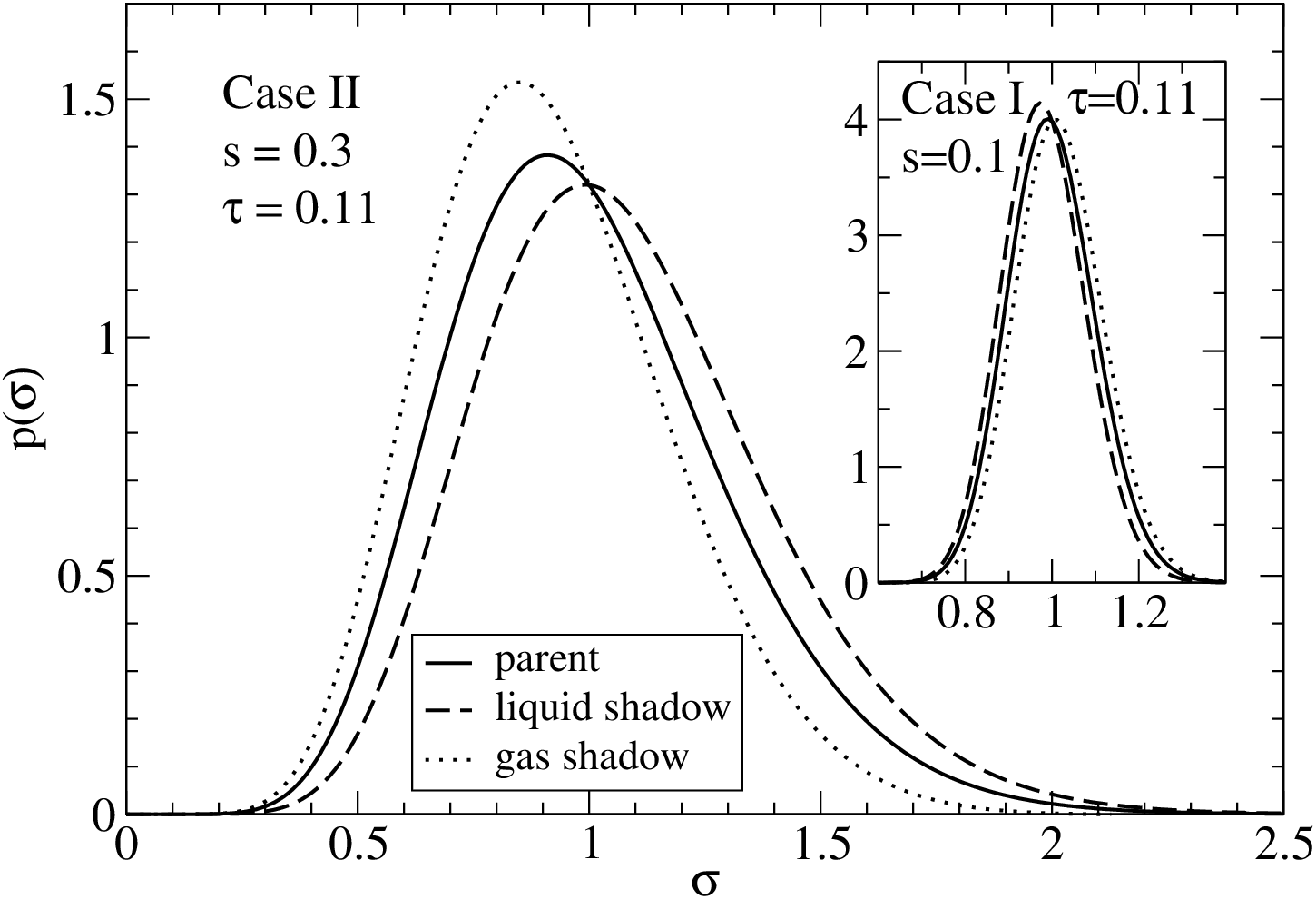}
\end{center}
\caption{}
\label{fig:disp}
\end{figure}
%
%
\begin{figure}[ht!]
\begin{center}
\includegraphics[width=14cm]{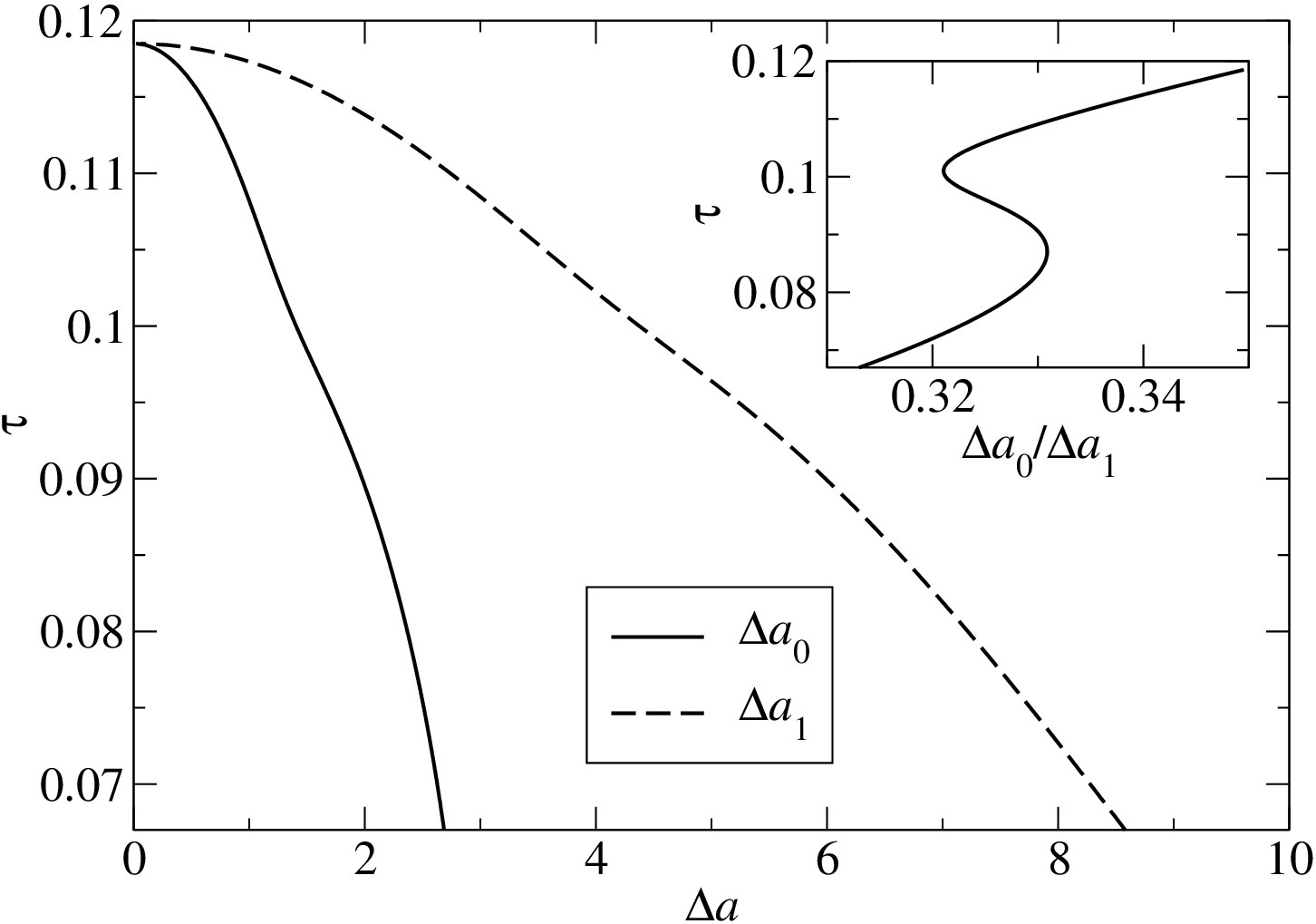}
\end{center}
\caption{}
\label{fig:delta_a}
\end{figure}
%
%
\begin{figure}[ht!]
\begin{center}
\includegraphics[width=14cm]{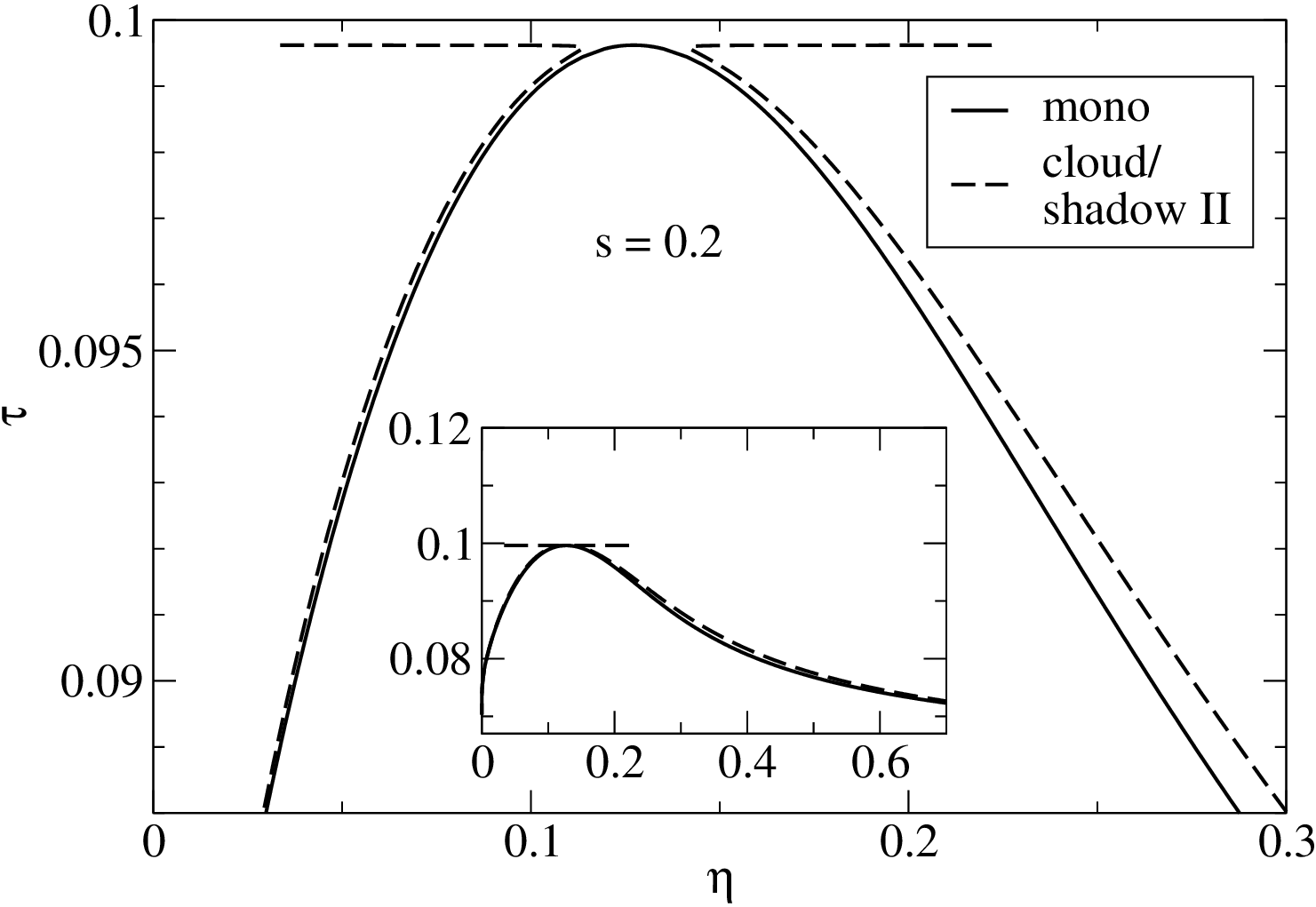}
\end{center}
\caption{}
\label{fig:csb3}
\end{figure}
%
%
\begin{figure}[ht!]
\begin{center}
\includegraphics[width=14cm]{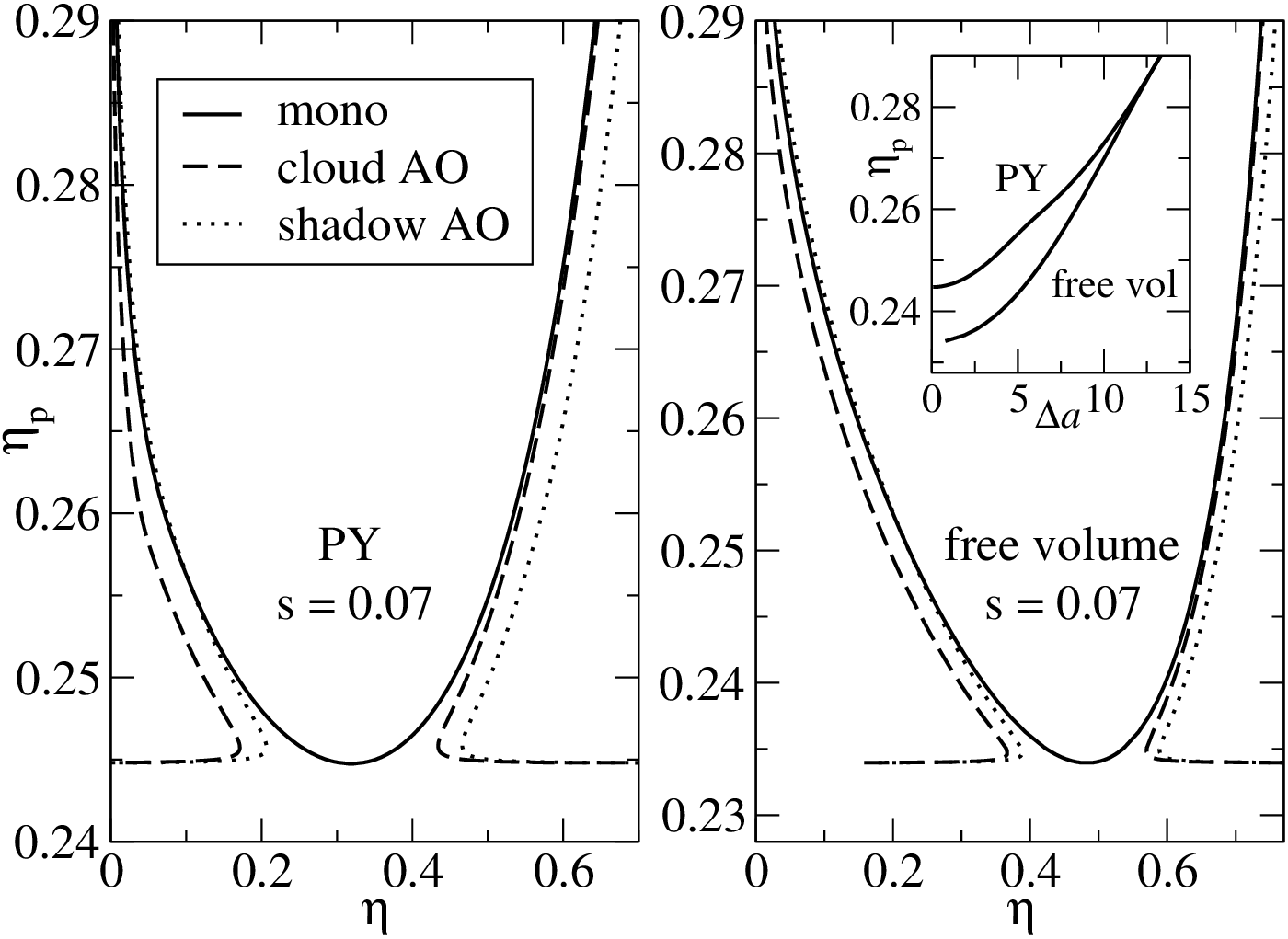}
\end{center}
\caption{}
\label{fig:free_vol}
\end{figure}

\end{document}